\documentclass[fleqn,usenatbib]{mnras}

\usepackage{newtxtext,newtxmath}

\usepackage[T1]{fontenc}
\usepackage{threeparttable}
\usepackage{ulem}
\DeclareRobustCommand{\VAN}[3]{#2}
\let\VANthebibliography\thebibliography
\def\thebibliography{\DeclareRobustCommand{\VAN}[3]{##3}\VANthebibliography}

\usepackage{graphicx}	
\usepackage{amsmath}	

\usepackage{amssymb}	
\usepackage{multirow}
\usepackage{booktabs}
\usepackage{color}
\usepackage{bm}
\usepackage{adjustbox}
\usepackage{tablefootnote}

\title[Detecting EM counterparts from lensed BNS merger GW events]{On the detection of the electromagnetic counterparts from lensed gravitational wave events by binary neutron star mergers}
\author[Ma et al.]{
Hao Ma,
Youjun Lu\thanks{luyj@nao.cas.cn},
Xiao Guo,
Siqi Zhang,
and 
Qingbo Chu
\\
$^{1}$\,National Astronomical Observatories, Chinese Academy of Sciences, 20A Datun Road, Beijing 100101, China\\
$^{2}$\,School of Astronomy and Space Sciences, University of Chinese Academy of Sciences, 19A Yuquan Road, Beijing 100049, China
}

\date{Accepted XXX. Received YYY; in original form ZZZ}

\pubyear{2022}

\begin{document}
\label{firstpage}
\pagerange{\pageref{firstpage}--\pageref{lastpage}}
\maketitle

\begin{abstract}
Future ground-based gravitational wave (GW) detectors, i.e., Einstein telescope (ET) and Cosmic Explorer (CE), are expected to detect a significant number of lensed binary neutron star (BNS) mergers, which may provide a unique tool to probe cosmology. In this paper, we investigate the detectability of the optical/infrared electromagnetic (EM) counterparts (kilonovae/afterglows) from these lensed BNS mergers by future GW detectors and EM telescopes using simple kilonova, afterglow, and lens models. ET and CE are expected to detect $\sim5.32^{+26.1}_{-5.10}$ and $67.3^{+332}_{-64.7}$ lensed BNS mergers per year. We find that the EM counterparts associated with all these mergers will be detectable by an all sky-survey in the H-band with the limiting magnitude $m_{\textrm{lim}}\gtrsim27$, while the detectable fraction is $\lesssim0.4\%$ in the g-/z-band if with $m_{\textrm{lim}}\lesssim24$. Generally it is more efficient to search the lensed EM counterparts by adopting the infrared bands than the optical/UV bands with the same $m_{\textrm{lim}}$. Future telescopes like Vera C. Rubin Observatory, China Space Station Telescope, and Euclid can hardly detect the EM counterparts of even one lensed BNS merger. Roman Space Telescope (RST) and James Webb Space Telescope (JWST) have the capability to detect about a few or more such events per year. Moreover, the time delays and separations between the lensed image pairs are typically in the ranges from minutes to months and from $0.1$ to $1$\,arcsec, suggesting that both the GW and EM images of most lensed BNS mergers can be well resolved by not only CE/ET in the time domain but also RST/JWST spatially.
\end{abstract}

\begin{keywords}
gravitational lensing: strong - gravitational waves - (stars:) gamma ray bursts: general - (transients:) neutron star mergers
\end{keywords}



\section{Introduction}
\label{sec:intro}

Gravitational wave and multi-messenger astronomy is blossomed after the detection of the binary black hole (BBH) merger GW150914 \citep{LIGO2016gw150914} and binary neutron star (BNS) merger GW170817 \citep{LIGO2017a}. The laser interferometer gravitational wave observatories (LIGO) and Virgo have already detected at least two BNSs, three neutron star-black hole binaries, and more than eighty BBHs \citep{GWTC3pop,LIGONSBH2021}. With further upgrade, the Advanced LIGO plus \citep[LIGO A+;][]{LIGO2020LVK} and LIGO Voyager \citep[Voyager;][]{Adhikari=2020} are expected to detect many more in the near future. The third generation GW detectors, i.e., the Einstein Telescope \citep[ET;][]{Punturo=2010,Hild=2011} and the Cosmic Explorer \citep[CE;][]{Abbott=2017,Reitze=2019CE} are expected to detect more than several tens of thousands of BBH and BNS mergers per year, let alone the proposal of moon-based Gravitational-wave Lunar Observatory for Cosmology \citep[GLOC;][]{Jani=2020}. Among these GW events, about a fraction $\sim 10^{-3} - 10^{-4}$ of them are expected to be gravitational lensed \citep{Oguri=2010, Biesiada=2014, LiSS=2018}. Searching and identifying the lensed GW events becomes one of the main goals of GW detection, as they not only are interesting phenomena but also can be used as a unique tool to study the nature of GW \citep[e.g.,][]{Mukherjee=2020} and probe cosmology \citep[e.g.,][]{Liao=2017, Hou=2020, Urrutia=2021arXiv,Guo=2022}. 

It has been proposed that the lensed GW events with independently measured redshifts can be applied to constrain Hubble constant accurately encoded in the ``time delay distance'' as the time delay can be precisely measured from the GW observations \citep{Treu=2016,Liao=2017,Hou=2020, Cremonese=2020, Hannuksela=2020}. For example, \citet{Liao=2017} have shown that the Hubble constant can be constrained to a level of $\lesssim 1\%$ precision by using only $10$ lensed BNS mergers with known redshifts, and \citet{Hannuksela=2020} also demonstrated such an application with detailed modelling of the lens systems. 

It is possible to get indirect redshift estimates via the detection of their lensed host galaxies and tightly constrain the Hubble constant \citep[see][for the mergers of stellar BBHs]{YuH=2020,Hannuksela=2020,Wempe=2022arXiv}. However, other cosmological parameters, such as the fraction of dark matter and dark energy, cannot be well constrained due to the degeneracy between the true luminosity distances and the magnification factors.
If the exact locations of these lensed GW events in their hosts can be determined, the magnification factor of each lensed images can be obtained by the reconstruction of the lens and thus the true luminosity distance can be obtained. In such a case, the time delay distance and the luminosity distance can be applied simultaneously, which enables tighter constraints on other cosmological parameters \citep{LIGO_siren=2017,Chen_siren=2018,Hannuksela=2020}.

One may measure the exact locations and also the redshifts of the lensed GW events by directly detect their lensed EM counterparts. This is possible for those compact binary mergers with a neutron star component, which lead to EM counterparts like kilonovae \citep[e.g.,][]{LiLX=1998, Metzger2019} and afterglow signals from short-duration gamma-ray burst (SGRB) \citep[e.g.,][]{Metzger=2012, Alexander=2017}. These EM signals, if detected, could provide much more precise localization of the lensed events and thus break the degeneracy in measuring cosmological parameters by using time delays. However, most previous studies are focused on estimating the detection rate of the lensed GW events but seldom consider the detectability of their EM counterparts \citep{Piorkowska=2013,Biesiada=2014, LiSS=2018, Oguri=2018, Wang=2021, Yang=2022, Xu=2022}. In this paper, we investigate the detectability of the EM counterparts of the lensed BNS mergers detected by GW detectors for different limiting magnitude conditions, mainly focusing on the kilonovae and SGRB afterglows. Furthermore, we estimate the detection rate of those events of which both the GW and the EM signals can be detected, by future telescopes, such as the China Space Station Telescope \citep[CSST;][]{Gong=2019}, the Vera C. Rubin Observatory \citep[Rubin;][]{Ivezic=2019}, the Nancy Grace Roman Space Telescope \citep[RST;][]{Dore=2019}, the Euclid \citep{Euclid2020}, and the James Webb Space Telescope (JWST)\footnote{\url{https://www.jwst.nasa.gov/}}. 

The paper is organized as follows. In Section~\ref{sec:method}, we introduce the methodology for quantitative estimates of gravitational lensed GW events and their associated kilonova and afterglow phenomena. Our results are presented in Section~\ref{sec:results}. Conclusions and discussions are given in Section~\ref{sec:conclusions}. 
Throughout this paper, we adopt the concordance $\Lambda$CDM cosmology model with $(\Omega_{\rm m}, \Omega_{\Lambda})=(0.3,0.7)$ and the Hubble constant $H_0= 70.0\ {\rm km\,s^{-1}\,Mpc^{-1}}$.

\section{METHODOLOGY}
\label{sec:method} 

In this section, we first introduce the lens statistics by adopting the singular isothermal ellipsoid (SIE) lens model with external shear (Section~\ref{sec:lensstat}) \citep{Schneider=2006, Oguri=2010, Oguri=2018, LiSS=2018}. Then, we describe our method to estimate the detection rate for the lensed BNS merger GW events (Section~\ref{sec:gwlens}), their associated kilonovae and afterglows signals (Section~\ref{sec:EMlens}), and both the GW events and the EM signals (Section~\ref{subsec:GWpEM}). Note here that we only consider the strong gravitational lensing events in the geometrical optics limit as the diffraction effect can be safely ignored in the LIGO band ($10 \sim 10^{3} \rm{Hz}$) for galaxy lenses \citep{Takahashi=2003}.

\subsection{Lensing Statistics}
\label{sec:lensstat}
 
We denote the GW event rate detected by a GW detector as $\Phi(\varrho,z_{\rm s})$ with the signal-to-noise ratio (SNR) in the range from $\varrho$ to $\varrho+d\varrho$ and redshift from $z_{\rm s}$ to $z_{\rm s}+dz_{\rm s}$. For simplicity, we assume all the BNS mergers have two types of EM counterparts kilonovae and afterglows. Thus event rate considering GW+EM joint detection is denoted as $\Psi(\varrho,m_{\rm _{EM}},z_{\rm s})$ similar to $\Phi(\varrho,z_{\rm s})$, where $m_{\rm{_{EM}}}$ is the apparent magnitude for individual EM counterpart. The strong lensing is mainly caused by the intervening early-type galaxies \citep{Turner=1984,Moller=2007}, of which the number density distribution can be denoted as $dN/d\sigma_v$ within the velocity dispersion range from $\sigma_v$ to $\sigma_v+d\sigma_v$. We adopt the power-law evolution model to describe the redshift evolution of the distribution of $dN/d\sigma_v$ described by the modified Schechter function \citep{Oguri=2012, Geng=2021} 
\begin{equation}
\frac{\mathrm{d} N}{\mathrm{~d} \sigma_{\rm v}}=\phi_{z} \left(\frac{\sigma_{\rm v}}{\sigma_{z} }\right)^{\alpha} \exp \left[-\left(\frac{\sigma_{\rm v}}{\sigma_{z}}\right)^{\beta}\right] \frac{\beta}{\Gamma(\alpha / \beta)} \frac{1}{\sigma_{\rm v}}
\end{equation}
and
\begin{equation}
\phi_{z} = \phi_{*} \left( 1+z_{\rm l} \right)^{\kappa_n};\ \ \sigma_{z} = \sigma_{*} \left( 1+z_{\rm l} \right)^{\kappa_{\rm v}},
\end{equation}
where ($\phi_{*}$, $\sigma_{*}$, $\alpha$, $\beta$)=($8.0 \times 10^{-3}h^{3} {\rm Mpc}^{-3}$, $161\ {\rm km\ s^{-1}}$, $2.32$, $2.67$) are the fitting results given by \citet{Choi=2007}, and two redshift evolution parameters $\kappa_{n}=-1.18$ and $\kappa_{v}=0.18$ are from \citet{Geng=2021}.

In the geometrical optics regime, the probability that a GW event and its EM counterparts are lensed can be characterised by the \textit{optical depth}. Similar to \citet[][see also  \citealt{Huterer=2005} and \citealt{Schneider=2006}]{LiSS=2018}, the optical depth for a GW event detected by a GW detector with SNR of $\varrho$ at redshift $z_{\rm s}$ can be estimated as
\begin{eqnarray}
\label{eq:tau_GW}
& \tau_{{_{\rm GW}}}\left(\varrho, z_{\rm s}\right) = \frac{1}{4 \pi} \int_{0}^{z_{\rm s}} d V \int d \sigma_{\rm v}  \frac{d N}{d \sigma_{\rm v}}  \int d q  p(q) \times \nonumber \\
&  \iint d \boldsymbol{\gamma}  p\left(\gamma, \theta_{\gamma}\right) \int d \mu A \frac{p(\mu)}{\sqrt{\mu}} \frac{\Phi\left(\varrho / \sqrt{\mu}, z_{\rm s}\right)}{\Phi\left(\varrho, z_{\rm s}\right)}. 
\end{eqnarray}
Similarly at any given band, the optical depth for a GW event and its EM counterparts detected by both a GW detector with SNR of $\varrho$ and an EM telescope with an apparent magnitude of $m_{\rm _{EM}}$ at redshift $z_{\rm s}$ is given by 
\begin{eqnarray}
\label{eq:tau_GWEM}
& \tau_{\rm _{GW+EM}} (\varrho,m_{\rm _{EM}},z_{\rm s}) =
\frac{1}{4 \pi} \int_{0}^{z_{\rm s}} d V \int d \sigma_{v}  \frac{d N}{d \sigma_{v}}  \int dq  p(q)  \times \nonumber \\ 
&  \iint d \boldsymbol{\gamma}  p\left(\gamma, \theta_{\gamma}\right) \int d \mu A \frac{p(\mu)}{\sqrt{\mu}} \frac{\Psi(\varrho/\sqrt{\mu},m_{\rm _{EM}}+2.5 \log \mu,z_{\rm s}))}{\Psi(\varrho,m_{\rm _{EM}},z_{\rm s})}. 
\end{eqnarray}
In the above two equations, $d V$ is the comoving volume element with redshift in the range $z \rightarrow z + dz $, $p(q)$ and $p\left(\gamma,\theta_{\gamma}\right)$ represent the probability distributions of the axis ratio $q$ and the two dimensional external shear ($\gamma,\theta_\gamma$), which describe the lens morphology and external environment near the line of sight \citep{LiSS=2018}, $\mu$ represents the magnification of the lensed images, with which the GW SNR is magnified by a factor $\sqrt{\mu}$, and the EM counterparts magnitude is brightened by $2.5 \log \mu$ \citep{Schneider=2006,Oguri=2018,LiSS=2018}, and $p(\mu)$ represents the probability distribution of $\mu$, $A$ represents the cross-section of the lens determined by the assembles of source locations having multiple lensed images. The cross-section is related to the lens galaxy velocity dispersion, the redshifts of lens and source, axis ratio and external shear. It is more convenient to consider the dimensionless version of cross-section $\tilde{A}(q,\boldsymbol{\gamma})=A/\theta_{E}^{2}$, with which the parameters of lens velocity dispersion and lens and source redshift are separated out into the angular Einstein radius given by $\theta_{E} = 4 \pi (\sigma_{\rm v}/c)^{2} (D_{\rm ls}/D_{\rm s})$. $D_{\rm ls}$ and $D_{\rm s}$ denote the angular diameter distance between the lens and the source and between the observer and the source.

\subsection{Models for gravitational waves}
\label{sec:gwlens}

In the case of GW, we consider the most general model which is the Newtonian approximation described by quadrupolar formula. Under such an assumption, $\Phi\left(\varrho, z\right)$ is \citep{Finn1996,LiSS=2018}
\begin{equation}
\Phi\left(\varrho,z \right)=\int d \mathcal{M}_{0} \frac{d V}{d z} \frac{\mathcal{R}_{\rm mrg}\left(\mathcal{M}_{0} ; z\right)}{(1+z)} P_{\varrho}\left(\varrho | z, \mathcal{M}_{0}\right).
\end{equation}
Here $\mathcal{M}_{0}$ is the intrinsic chirp mass, $\mathcal{R}_{\rm mrg}(\mathcal{M}_0;z)$ is the merger rate density of GW source with intrinsic chirp mass $\mathcal{M}_0\rightarrow \mathcal{M}_0+ d\mathcal{M}_0$ at redshift $z$, and $P_{\varrho}\left(\varrho|z,\mathcal{M}_{0}\right)$ represents the probability distribution of a GW source with chirp mass $\mathcal{M}_0$ at redshift $z$ detected with a SNR of $\varrho$ as \citep{Finn1996} 
\begin{equation}
P_{\varrho}\left(\varrho | z ,\mathcal{M}_{0}\right) =\left.P_{\Theta}(\Theta(\varrho)) \frac{\partial \Theta}{\partial \varrho}\right|_{\mathcal{M}_{0}, z},
\end{equation}
and
\begin{equation}
\Theta(\varrho) = \frac{\varrho}{8} \frac{D_{\rm L}(z)}{R_{0}}\left(\frac{1.2 M_{\odot}}{\mathcal{M}_0(1+z)}\right)^{5 / 6} \frac{1}{\sqrt{\zeta (f_{\rm{max}})}}.
\end{equation}
In the above equations, $\Theta$ stands for the angular orientation function and it depends on the relative orientation of the detector and the source, $R_0$ is characteristic distance for a given GW detector which depends on the detector’s noise power spectral density $S_{n}(f)$, and it describes a detector’s detection capability, $D_{\rm L}(z)$ is the luminosity distance at redshift $z$, and $\zeta (f_{\rm{max}})$ represents the overlap between GW signal and effective bandwidth of detectors, where we adopt $\zeta (f_{\rm{max}})=1$ for simplicity \citep{LiSS=2018}. In this work, we consider five future GW detectors, i.e., LIGO A+, LIGO Voyager, ET, CE, and GLOC\footnote{Sensitivity curve data is from \url{https://dcc.ligo.org/LIGO-T1800042/public} for LIGO A+, \url{https://dcc-lho.ligo.org/LIGO-T1500293/public} for LIGO Voyager, \url{http://www.et-gw.eu/} for ET-D design \citep{Hild=2011}, \url{https://cosmicexplorer.org/} for CE Stage-2 phase \citep{Reitze=2019CE} (unless otherwise specified, CE in this work all refers to CE Stage-2), and \url{ https://doi.org/10.5281/zenodo.3948466} for GLOC.}, respectively. With the designed sensitivity curves of these detectors, we have $R_{0}= 194$\,Mpc, $477$\,Mpc, $1586$\,Mpc, $4034$\,Mpc, and $5688$\,Mpc, correspondingly. According to \citet{Finn1996}, $\Theta$ is a function of $\varrho$ and its probability distribution $P(\Theta)$ can be well approximated by a piecewise function, i.e., $P_{\Theta}(\Theta)=5 \Theta(4-\Theta)^{3} /256$ when $0<\Theta<4 $, and $P_{\Theta}(\Theta)=0$ otherwise.  

The intrinsic merger rate density $R_{\rm mrg}(\mathcal{M}_{0};z)$ can be estimated by combining the population synthesis models for binary stellar evolution (BSE) and the models of cosmological galaxy formation and evolution \citep[e.g.,][]{Chu=2022, Belczynski=2002,Dominik=2012,Dominik=2013,Dominik=2015,Yu=2015,Giacobbo=2018,Belczynski=2018}. We adopt the results on the BNS merger rate obtained by implementing the BSE model ``$\alpha 10.\rm{kb}\beta 0.9$'' into the galaxy formation and evolution model Millennium-II in \citet{Chu=2022}, which can well match the distribution of the observed BNSs in the Milky Way. We re-scale this predicted merger rate density evolution to make it the same as the local merger rate density determined by the LIGO/Virgo observations, i.e., $320\,{\rm Gpc^{-3}yr^{-1}}$ \citep{GWTC2}. The latest constraint from GWTC-3 \citep{GWTC3pop} shows the local BNS merger rate should be in the range from $13$ to $1900\,{\rm Gpc^{-3}yr^{-1}}$, and we take these numbers as the indicators of the uncertainties in the estimate of the local BNS merger rate density, and correspondingly it leads to an uncertainty in our estimates for the GW and EM detection rates of the BNS mergers. Note here that the general trend for the merger rate density evolution may be also different in different models \citep[e.g.,][]{Mandel=2022}, which may slightly affect our estimates below for the detection of lensed GW and EM signals. For simplicity, however, we ignore this uncertainty and take it as a secondary effect.

\subsection{Kilonovae and afterglows associated with BNS mergers}
\label{sec:EMlens}

It is already known that kilonovae are promising EM counterparts of BNS mergers and can be detected by ground-based or spaceborne telescopes \citep{LIGO2017EM}. Comparing with the beamed GRB and its afterglow (see later description in this subsection), kilonova powered by $r$-process is relatively isotropic and thus may be the most promising detectable EM counterpart of the BNS merger \citep{Metzger2019}. The light curves (LCs) of the BNS merger GW170817 can be well fit by a two-component or three-component kilonova model \citep[e.g.,][]{Cowperthwaite=2017,Villar=2017}. More complicated kilonova models also proposed to interpret the detailed multiband observations of GW170817 \citep[e.g.,][]{Margutti=2021, Nicholl=2021, Just=2022}. In this paper, we adopt the isotropic two-component kilonova model given in \citet{Villar=2017} to generate the LCs at different bands for simplicity. This model assumes only the `red' and `blue' components in kilonova emissions, in which `red' and `blue' are determined by the abundance of lanthanide. The `red' component corresponds to lanthanide-rich ejecta with a relatively higher opacity, while the `blue' component corresponds to lanthanide-poor ejecta with a lower opacity. The emissions from the `red' and `blue' components peak at relatively redder and bluer bands, respectively. Each of the component is mainly described by four parameters, i.e., ejecta mass ($M_{\rm ej}$), ejecta velocity ($v_{\rm ej}$), opacity ($\kappa_{\rm ej}$), and the temperature floor ($T_{\rm f}$).  

\begin{figure}
\centering
\includegraphics[width=0.45\textwidth]{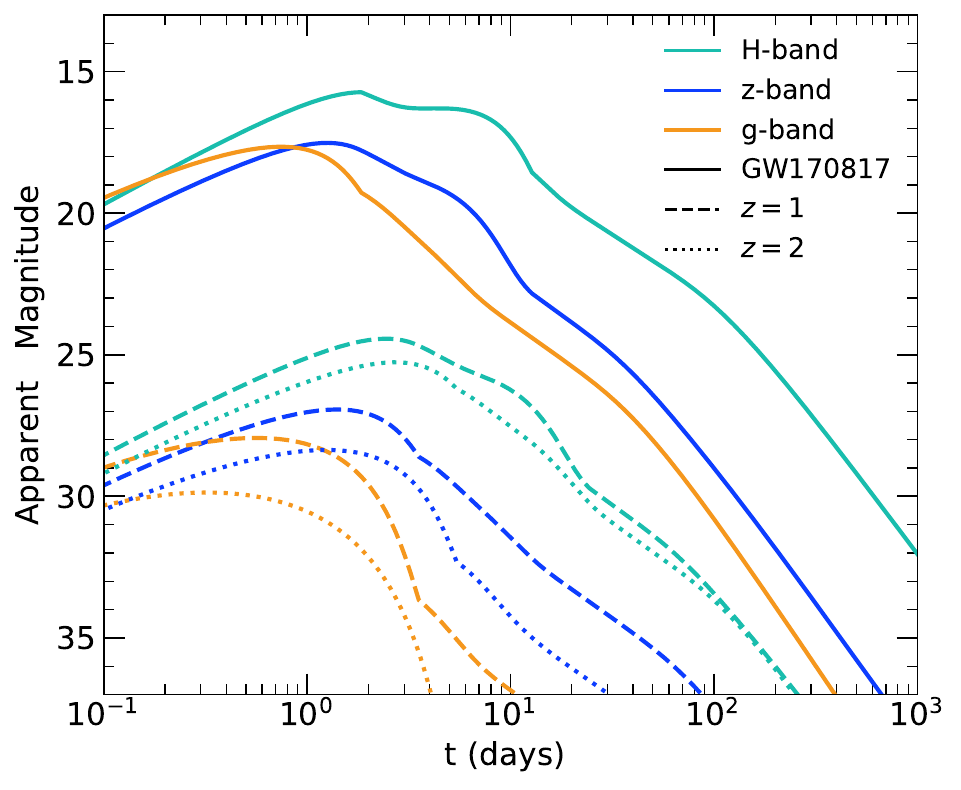}
\caption{Example light curves for kilonovae due to BNS mergers as a function of the observation time $t$. Here $t=0$ represents the merger time. Solid lines represent the simulated light curves of GW170817, and the dashed and dotted lines represent the light curves of kilonovae the same as GW170817 but located at $z=1$ and $z=2$, respectively. Cyan, blue, and orange lines represent the light curves of H-, z-, and g-bands, respectively.
}
\label{fig:f1}
\end{figure}

Figure~\ref{fig:f1} shows the H-, z-, and g-band\footnote{Unless otherwise specified, the H-band is chosen from the Euclid, z- and g-bands are chosen from CSST in this work. All the filter data are provided by the Spanish Virtual Observatory (SVO) Filter Profile Service \citep{SVO1, SVO2}.} LCs of a kilonova like GW170817 located at several different redshifts, obtained by adopting the best-fit two-component model given in \citet[][see their Tab. 2 for model parameters]{Villar=2017}. The shape of the observed LCs of a kilonova, same as GW170817 but locating at redshift $z=1$ or $2$, are different from those of GW170817 because of the time dilation and the K-correction. We define the peak time $T_{\rm p}$ as the time for a kilonova at any specific band reaching its peak magnitude after the merger, and the peak duration of that LC $\Delta T_{\rm p-0.5}$ as the duration of the kilonovae LCs with magnitude from the peak magnitude $m_{\rm p}$ to $m_{\rm p}+0.5$. Apparently, $T_{\rm p}$ and $\Delta T_{\rm p-0.5}$ change with redshift as a combined effect of the time dilation and the K-correction.

\begin{figure}
\centering
\includegraphics[width=0.46\textwidth]{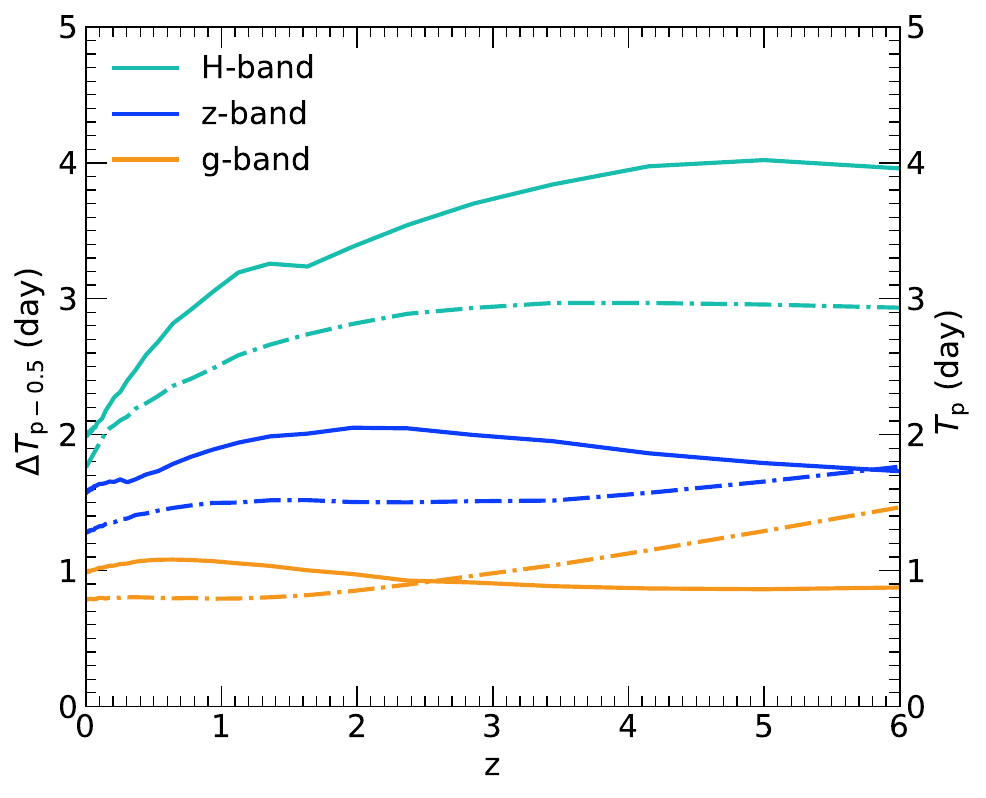}
\caption{
The time interval $\Delta T_{\rm p-0.5}$ (solid lines) and the peak time $T_{\rm p}$ (dotted-dashed lines) for the LCs of kilnovae at different redshift ($z$). Here $\Delta T_{\rm p-0.5}$ is defined as the period when the kilonova magnitude at any specific band is in the range from $m_{\rm p}+0.5$ to $m_{\rm p}$ with $m_{\rm p}$ denoting the peak magnitude, and $T_{\rm p}$ denotes the time that the kilonova reaches the peak magnitude at that specific band after the merger. Cyan, blue, and orange lines show the results for the H-, z-, and g-band, respectively.
}
\label{fig:f2}
\end{figure}

Figure~\ref{fig:f2} shows $T_{\rm p}$  and $\Delta T_{\rm p-0.5}$ for the LCs of kilonovae at different redshift. For illustration, we only show those for the H-, z-, and g-band LCs, respectively. As seen from this figure, $T_{\rm p}  \gtrsim 1\, \rm{day}$ and $\Delta T_{\rm p-0.5} \gtrsim 1\, \rm{day}$ for all the LCs at H-, z-, and g-bands, and the LCs in the redder bands have relatively larger $T_{\rm p}$ and larger $\Delta T_{\rm p-0.5}$ comparing with those in the bluer bands. These suggest that the time should be sufficient for telescopes to prepare for searching the appearance of kilonovae of BNS mergers at its peak luminosities. Adopting a much shorter response time ( $\ll 1$\,day) may not significantly improve the detectability of the kilonovae, the main concern of this paper\footnote{One should note here that a shorter response time is extremely important for studying the underlying physical processes of kilonova and the nature of neutron stars, since it can be applied on distinguishing different models for interpreting the early blue emission of kilonovae \citep[e.g. radioactive decay in low-opacity ejecta, relativistic boosting of radioactive decay in high-velocity ejecta, and the cooling of the material heated by a wind/jet. See][]{Arcavi=2018}.}.

In addition, due to the cosmological redshift effect, the searching of distant kilonovae ($z \gtrsim 1$) which is the majority of the lensed BNSs (see Fig.~\ref{fig:f7}), can be more efficient by using the redder filters than the optical filters. Note here that for those kilonovae in the local universe, the optical filters have unparalleled advantages in observing them, since the infrared and near-infrared filters have a much lower throughput and multi-optical-band observations can be used to separate kilonovae photometrically from other transients \citep{Bianco=2019,Andreoni=2019,Andreoni=2022}. In principle, one may design the search strategy of each individual kilonovae according to its expected LCs. Without loss of generality, we assume that the kilonova can always be searched for during the period with magnitude no fainter than the peak magnitudes by $0.5$\,mag. 

For the afterglow, we assume a simple Gaussian jet model by using the package \texttt{afterglowpy}, developed by \citet{Ryan=2020}, to estimate the LCs. According to \citet{Ghirlanda=2013}, the jet opening angle $\theta_{j}$ (in unit of degree) of the afterglow follows a log-normal distribution $\propto (1 /\sigma \theta_j) \exp\left[-(\ln \theta_j-\mu)^2/ (2\sigma^2)\right]$, with $\mu=1.742$ and $\sigma=0.916$. Note that the afterglow of GW170817 can also be well reconstructed by the Gaussian jet model and the best-fit value for $\theta_j$ is $3.27^\circ$, consistent with the log-normal distribution given by \citet{Ghirlanda=2013}. For observers viewing the afterglow at off-axis, the received afterglow emission drops rapidly when the viewing angle $\theta_{\rm obs}$ increase to larger than a truncation angle $\theta_{w}$, and the kilonova radiation normally dominates when $\theta_{\rm obs}>\theta_{w}$. In our calculations, we fix $\theta_{w}=2\theta_{j}$ according to \citet{Metzger=2012}. Other parameters are fixed to the best-fit values obtained by fitting  the late-time broad-band EM observations of GW170817 to the Gaussian jet model \citep{Troja=2018}. These parameters are the on-axis equivalent isotropic energy $\log E_{0}/{\rm erg}=52.73$, the number density of the interstellar medium $n_{0}=10^{-3.8} {\rm cm}^{-3}$, the power-law index of accelerated shock $p=2.155$, the magnetic field energy fraction $\epsilon_{e}=10^{-1.51}$ and the accelerated electron energy fraction $\epsilon_{B}=10^{-3.20}$. 

\begin{figure}
\centering
\includegraphics[width=0.45\textwidth]{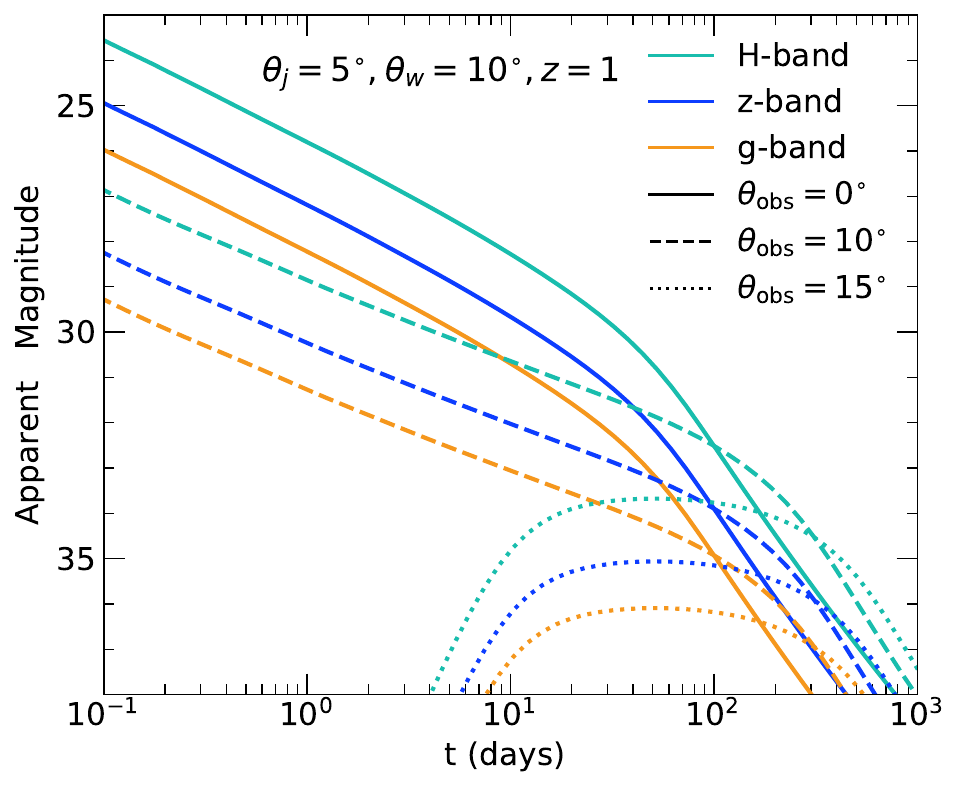}
\caption{Example light curves for afterglows of sGRBs due to BNS mergers obtained by using the package \texttt{afterglowpy} \citep{Ryan=2020}. For illustrative purposes, the jet opening angle $\theta_j$ is fixed at $5^\circ$ and the redshift $z$ is fixed at $1$ here. The truncation angle $\theta_{\rm w}$ is assumed to be $2\theta_j$ \citep[e.g., see][]{Metzger=2012}. Other afterglow model parameters are in line with those introduced in Section~\ref{sec:gwlens}. Solid, dashed and dotted lines represent those obtained with viewing angles of $\theta_{\rm obs}=0^\circ$, $10^\circ$, and $15^\circ$ (from on-axis to off-axis angle $3\theta_{j}$), respectively. Cyan, blue, and orange lines represent the light curves of H-band, z-band, and g-band respectively.
}
\label{fig:f3}
\end{figure}

Figure~\ref{fig:f3} illustrates some example cases of the afterglow LCs at the H-, z-, and g-band respectively, in which time dilation and K-correction are taking into consideration. As seen from this figure, the afterglow magnitudes decrease with elapsing time $t$ at all the three bands. For the bright cases with small viewing angle, the luminosities decrease rapidly monotonically at small $t$, which suggests that a short response time (i.e., the time between the searching of the afterglow and the GW alert of the BNS merger) is critical for searching afterglows. It is easier to detect the afterglow if the response time is shorter, comparing with the kilonova case in which a short response time ($\ll 1$\,day) of observation is corresponding to fainter magnitude since the peaks of kilonova LCs normally emerge at $T_{\rm p} \gtrsim 1\, \rm{day}$ (see Figure~\ref{fig:f2}). 

The recent transient searches triggered by GW alerts have reported a mean response time of $9.90$\,hr \citep{Gompertz=2020} and of $1.5$\,hr \citep{Kasliwal=2020} during the first half of the LVC O3 run. Thus, we assume two response time cases of $10$\,hr and $1.0$\,hr in the following calculations, denoting the typical time period between the detection of the GW from the BNS merger and the EM counterparts searching time, and calculate the afterglow magnitude at this time for different bands. Note again that the detectability of the EM counterparts is the main concern of this paper. Adopting a much short or long response time may not significantly improve the detectability of kilonova (see Fig.~\ref{fig:f6}) as the peak luminosities emerge at 1\,day $\lesssim T_{\rm p} \lesssim 4$\,day (see Fig.~\ref{fig:f2}), though the observations in the first few hours and/or the UV band are indeed important for constraining the kilonova model and the nature of neutron stars. Therefore, we assume that the searching observations for kilonovae can always be performed in a period around the peaks of the light curves for a response time of $10$\,hr. We also set the magnitude of both kilonovae and afterglows at $1.0$\,hr for the case with response time of $1.0$\,hr and compare those results obtained from this case with the above one.
 
The luminosity and its evolution at any given band of an afterglow is sensitive to the viewing angle $\theta_{\rm obs}$ of the binaries (see Figure~\ref{fig:f3}), which is different from the nearly isotropic kilonova emission. The afterglow magnitudes decrease sharply with increasing viewing angle $\theta_{\rm obs}$, especially at $\theta_{\rm obs} > \theta_{\rm w}$, if the elapse time is less than $10$\,days since the merger. Note that the BNS mergers detected by the GW detectors have the viewing angle distribution of $P_{\rm det}(\theta_{\rm obs}) = 0.076(1 + 6 \cos^{2}{\theta_{\rm obs}} + \cos^{4}{\theta_{\rm obs}})^{3/2} \sin{\theta_{\rm obs}}$, due to the GW radiation pattern \citep{Schutz=2011, Metzger=2012}. Therefore, we set the view angles of afterglows following this probability distribution for the calculations of the mock BNS mergers in Section~\ref{sec:results}. 

\subsection{Joint detection of lensed GW+EM events}
\label{subsec:GWpEM}

For detection of the EM counterparts of GW events, i.e., kilonovae and afterglows, for demonstration, we consider some future ground-based/spaceborne telescopes, including Rubin, CSST, Euclid, RST, and JWST. All these optical/infrared telescopes will have their first observations in this decade and may be still the premier observatories of the next decade. For each telescope we pick one specific bandpass filter, which is the z filter of Rubin, the z filter of CSST, the H filter of Euclid, the filter $\rm{H158}$ of RST, and the filter $\rm{F150W}$ of JWST. We note here that all these future telescopes are loaded more than one bandpass filters that cover a wide-range of wavelength. For instance, JWST NIRCam offers $29$ bandpass filters in the wavelength range $0.6-5.0\mu{\rm m}$. Our results below only represent the cases of the adopted specific bandpass filters. For future realistic EM counterpart searches for BNS mergers, it depends on which bandpass filters that are finally get involved. In addition, we also consider the K-correction for all the EM counterparts at any given redshift. Since the lensed events mainly peak around $z_{\rm s} \sim 1-2$, the K-correction leads to the shifts of the intrinsic optical band emission of kilonova (probably the peak of the spectral energy distribution) to redder bands. The red band luminosity of afterglow is also higher than the blue band one (see Figure~\ref{fig:f3}). Therefore, adopting the infrared filters may be more efficient in searching for the lensed EM counterparts (see Section~\ref{sec:rates}). This is the primary reason that we adopt the five bandpass filters above in the near-infrared band.

From the above GW and EM models, the total event rate of BNS mergers with the GW SNR $\varrho$ and the apparent magnitude of the kilonovae $m_{\rm KN}$ or the afterglow $m_{\rm AG}$ can be generated by
\begin{equation}
\label{eq:Psi_KN}
\Psi(\varrho,m_{\rm KN},z_{\rm s}) = 
\int\Phi(\varrho, z_{\rm s}) p(m_{\rm KN} | \boldsymbol{x}_{\rm KN})p(\boldsymbol{x}_{\rm KN})  d \boldsymbol{x}_{\rm KN}, 
\end{equation}
for kilonovae, or
\begin{equation}
\label{eq:Psi_AG}
\Psi(\varrho,m_{\rm AG},z_{\rm s}) = 
\int \Phi(\varrho, z_{\rm s}) p(m_{\rm AG} | \boldsymbol{x}_{\rm AG})p(\boldsymbol{x}_{\rm AG}) d \boldsymbol{x}_{\rm AG}, 
\end{equation}
for afterglow. Here $ p(m_{\rm KN} | \boldsymbol{x}_{\rm KN}) $ and $ p(m_{\rm AG} | \boldsymbol{x}_{\rm AG}) $ are the probability distributions of kilonova and afterglow apparent magnitudes at an observation time $t_{\rm obs}$, in which $\boldsymbol{x}_{\rm KN}=\{M_{\rm ej}^{\rm blue},v_{\rm ej}^{\rm blue},\kappa_{\rm ej}^{\rm blue},T_{\rm f}^{\rm blue},M_{\rm ej}^{\rm red},v_{\rm ej}^{\rm red},\kappa_{\rm ej}^{\rm red},T_{\rm f}^{\rm red},z_{\rm s},t_{\rm obs}\}$ and $\boldsymbol{x}_{\rm AG}=\{\theta_{\rm obs}, \theta_{j},E_0,n_0,p,\epsilon_e,\epsilon_B,z_{\rm s},t_{\rm obs}\}$ are the model parameters of kilonova and afterglow. The superscripts `red' and `blue' denote the red- and blue-components of the kilonova ejecta, respectively. In principle, one could obtain these two probability distributions by adopting detailed kilonova and afterglow models, by considering the dynamics, radiation processes, geometric configuration, and environment of each BNS merger with known physical properties. For simplicity, we assume in this paper that all the kilonovae have the same $\boldsymbol{x}_{\rm KN}$ as that for GW170817, constrained by GW and multi-band EM observations as described above in Section~\ref{sec:EMlens}, and the kilonova radiation is close to isotropic. Therefore, $m_{\rm KN}$ is only a function of redshift due to the K-correction, and $p(m_{\rm KN} | \boldsymbol{x}_{\rm KN}) p(\boldsymbol{x}_{\rm KN})$ is taken as a delta function for kilonovae at any given redshift. We also assume that all afterglows have the same $\boldsymbol{x}_{\rm AG}$ obtained by late-time broad-band EM observations of GW170817 as described above, except that the viewing angles $\theta_{\rm obs}$ and jet opening angles $\theta_{j}$ are different. Therefore, $m_{\rm AG}$ is only a function of $\theta_{\rm obs}$, $\theta_{j}$, and redshift $z_{\rm s}$. Similarly, $p(m_{\rm AG} | \boldsymbol{x}_{\rm AG}) p(\boldsymbol{x}_{\rm AG})$ is also taken as a delta function at given $\theta_{\rm obs}$,$\theta_{j}$, and $z_{\rm s}$. In our following calculations, $\Psi(\varrho,m_{\rm KN},z_{\rm s})$ and $\Psi(\varrho,m_{\rm AG},z_{\rm s})$ are realized by sampling $m_{\rm KN} $ and $m_{\rm AG}$ using the kilonova and afterglow models described in Section~\ref{sec:EMlens} for sources at different redshift with different viewing angles and response time conditions. We assume that the afterglow signal is not correlated with the kilonova signal, and the kilonova is independent of the viewing angle $\theta_{\rm obs}$.

The lensed event rates for GW detection and GW+EM joint detection can be obtained by combining equations~\eqref{eq:tau_GW} and \eqref{eq:tau_GWEM}, i.e.,
\begin{eqnarray}
\frac{d \dot{N}_{\rm{L,GW}}}{d z_{\rm s}} & = & \int_{\varrho_{0}}^{\infty} d \varrho \  \Phi\left(\varrho, z_{\rm s}\right) \tau_{\rm{_{GW}}}  \nonumber \\ 
&= & g(z_{\rm s}) \int_{\varrho_{0}}^{\infty} d \varrho \int dq  p(q) \iint d \boldsymbol{\gamma}  p\left(\gamma, \theta_{\gamma}\right)  \nonumber \\
& & \times \iint \frac{d \boldsymbol{u}}{\sqrt{\mu}} \  \Phi\left(\varrho / \sqrt{\mu} , z_{\rm s}\right),
\label{eq:GW}
\end{eqnarray}
\begin{eqnarray}
\frac{d \dot{N}_{\rm L,GW+EM}}{d z_{\rm s}}& = & \int_{\varrho_{0}}^{\infty} d \varrho \int_{-\infty}^{m_{\rm{lim}}} d m_{\rm _{EM}} \Psi(\varrho,m_{\rm _{EM}},z_{\rm s})  \tau_{\rm GW+EM} \nonumber \\ 
&  = & g(z_{\rm s}) \int_{\varrho_{0}}^{\infty}  d \varrho \int_{-\infty}^{m_{\rm{lim}}} d m_{\rm _{EM}} \int dq  p(q) \iint d \boldsymbol{\gamma}  \nonumber  \\
&  & 
p\left(\gamma, \theta_{\gamma}\right) \iint \frac{d \boldsymbol{u}}{\sqrt{\mu}} \Psi(\varrho / \sqrt{\mu}, m_{\rm _{EM}}+2.5\log \mu ,z_{\rm s}). \nonumber \\
\label{eq:GWpEM}
\end{eqnarray}
In the above two equations,
\begin{equation}
g(z_{\rm s}) = \frac{1}{4 \pi} \int_{0}^{z_{\rm s}} d V \int d \sigma_{v}  \frac{d N}{d \sigma_{v}} \theta_{E}^{2}, 
\end{equation}
$\boldsymbol{u}$ is the angular position of the source in the source plane and is limited to within the cross-section region, the axis ratio $q$ follows a normal distribution truncated at $0.2$ and $1.0$ with a mean of $0.7$ and a standard deviation of $0.16$ \citep{Sheth=2003}, the external shear radial component $\gamma$ is well fitted with the logarithm normal distribution with mean of $0.05$ and standard deviation of $0.2 {\rm dex}$, and tangential component $\theta_{\gamma}$ is set to be random in the range from $0$ to $\pi$ \citep{Huterer=2005}. If set $\tau = 1$, the first lines of the above two equations give the intrinsic detectable event rates without considering the lensing effect. The EM counterpart magnitude $m_{\rm _{EM}}$ in equation~\eqref{eq:GWpEM} can be the magnitude of  either kilonova (`KN') or afterglow (`AG'). We also consider the rate for those cases that either KN or AG can be detected, i.e., 
\begin{equation}
\frac{d \dot{N}_{\rm{L,AG/KN}}}{d z_{\rm s}} = \frac{d \dot{N}_{\rm{L,GW+KN}}}{d z_{\rm s}}+\frac{d \dot{N}_{\rm{L,GW+AG}}}{d z_{\rm s}}-\frac{d \dot{N}_{\rm{L,GW+KN+AG}}}{d z_{\rm s}},
\label{eq:AGoKN}
\end{equation}
where the subscript `AG/KN' means either AG or KN is detected for those lensed BNS GW events,`GW+KN' and `GW+AG' represent that the KN and the AG signals are detected for the lensed GW events, respectively, and `GW+KN+AG' represents that all the GW, KN, and AG  signals can be detected for the same event. %

It is necessary to set the criteria for the detection of an lensed event with GW and EM signals for the estimation of those event rates. Different lensed images have different light paths and different arrival time. Once the first two arrived lensed GW signals of a BNS merger are detected, the event may be identified as a potential lensed system by analyzing their GW signals. Many recent works have assessed the feasibility of this pre-selection of lensed pairs by applying their sky localization, time delay, magnification ratio, and phase shift, etc \citep[e.g.,][]{Haris=2018, Hannuksela=2019, Dai=2020, LIGO2021lensing, Kim=2021,Liu=2021}. In the ET/CE era, the network of GW detectors can provide a medium localization accuracy around 10\,deg$^{2}$ \citep[e.g.,][]{Mills=2018, Zhao=2018, Chan=2018, LiYF=2022}. With this accuracy, random searches by general purpose or even some deep field sky survey telescopes may be not so efficient, due to the limitation of their Field of view (FoV). For example, the FoV of CSST, Euclid, RST, and JWST are $1.1$\,deg$^{2}$, $0.53$\,deg$^{2}$, $0.28$\,deg$^{2}$, and $9.7\,{\rm arcmin}^2$, respectively, which are more than tens to thousands times smaller than the typical localization area of GW events. It is almost impossible to use the general purpose telescopes, such as JWST, to scan the whole localization area to find the associated transients. However, the host galaxies of some lensed GW sources may be identified by matching the properties of the lensed GW events with those of the lensed hosts within the sky area \citep{YuH=2020,Wempe=2022arXiv}, with which the exact hosts for the EM counterparts can be obtained. Therefore, we ignore the FoV limitation and assume that the lensed host location of the lensed BNS merger events can be known at prior, thus both the deep field survey and general purpose telescopes can be considered. 

The pre-selection processes for the lensed GW events and its lensed host galaxies may lead to some uncertainties in the identification of the EM counterparts. For example, the processing time of the potential lensed system identification and its lensed host galaxy matching is a crucial factor and is supposed to be as short as possible. Other factors are the fraction of the lensed host galaxies that can be identifiable by the adopted survey telescope ($f_{\rm Host}$) and the success rate of the lensed host galaxy matching among all the candidates in the potential localization ($P_{\rm Host}$). For the general discussion in Section~\ref{sec:rates}, we assume that $f_{\rm Host}=100\%$ and $P_{\rm Host}=100\%$ at first. However, for the prediction of future specific telescopes, \citet{Wempe=2022arXiv} and \citet{ChenLu2022} estimated that the fraction of those lensed galaxies detected by sky surveys (e,g,, Euclid, CSST, etc.) is about $20-50\%$. Therefore, we adopt a moderate value of $f_{\rm Host}=30\%$ as this fraction in this paper. \citet{YuH=2020} found that the lensed GW signals and its host galaxy can be successfully matched within $10$\,deg$^{2}$ sky area for the 3rd generation GW detectors, while \citet{Wempe=2022arXiv} found the successful matching fraction could be as small as $\sim20\%$. Thus we adopt the success rate of matching $P_{\rm Host}=100\%$ as the optimistic case for later analysis if not otherwise stated. We also consider $P_{\rm Host}=20\%$ as the pessimistic case.

The arrival time sequence of the lensed images is important for the settings of the `detection criterion'. Assuming the SIE lens model, there will be two categories of lensed phenomenon, one with double images, and another with quadruple images. For the double-image cases, the second arrived images appear to be the fainter ones. Whereas for the quadruple-image cases, the first arrived images appear to be the third brightest ones, and the following two images are the first two brightest images \citep{Oguri=2010}. We fix the fainter images to be the second arrived ones for the double-image cases, while fix the brightest images as the second arrived ones for the quadruple-image cases. Although the accurate arrival time sequence can be uncertain and it depends on the configuration of the lens system, our assumption is appropriate in general. 
Below we consider two set of criteria for the identification of a lensed GW+EM event, i.e.,
\begin{itemize}
\item {\bf Regular criterion:} 1) The first two arrived GW signals of the BNS merger event can be detected, which means that the fainter image for the double-image cases or the third brightest ones for the quadruple-image case has the SNR above the detection threshold $\varrho_{0}$. This criterion is also consistent with the one frequently adopted in the statistical studies of GW lensed events  \citep[e.g.,][]{Oguri=2010, Oguri=2018, LiSS=2018, Yue=2022}; 2) The EM counterpart (either AG or KN) of the second arrived GW signal exceeds the detection magnitude threshold of the specific EM telescopes considered in this paper. With this criterion, at least one EM image of the lensed BNS merger event can be detected. 
    
\item {\bf Conservative criterion:} 1) The faintest GW signals for both the double-image and quadruple-image cases have the SNR above the detection threshold $\varrho_{0}$. 2) The faintest EM images for either the double-image or quadruple-image cases exceeds the detection magnitude threshold of the specific EM telescopes considered in this paper. Apparently, the conservative criterion is different from the regular criterion by only changing detection threshold for the quadruple-image cases. This means that the higher ratio of the cases with quadruple images to those with double images, the smaller detectable lensed event rate by adopting the conservative criterion than those estimated by adopting the regular criterion (see Table~\ref{tab:1}).

\end{itemize}

With the criterion for detectable lensed events defined above, the detectable rate of lensed events can be estimated according to equations~\eqref{eq:GW} and \eqref{eq:GWpEM}, which can be done by using the Monte Carlo method. Through solving the lens equation for each realized lens system, we can obtain detailed properties of their images, including the magnification factors, locations, arrival time, etc. \citep[][see also appendix in \citealt{LiSS=2018}]{Schneider=1992}.

\section{Results}
\label{sec:results} 

In this section, we present our results of lensed GW and its EM counterparts. We first predict the lensed GW event rate for five future GW detectors. For different limiting magnitude conditions, we calculate the detection efficiency (theoretical all sky lensed event rate), and also present the lensed GW+EM event rates of five specific upcoming telescopes (Section \ref{sec:rates}). Then redshift distributions and magnification distributions are demonstrated (Section~\ref{sec:redmag}). Finally from the perspective of time delays and angular separations of lensed images, we further explore the detection capabilities for future detectors and telescopes (Section~\ref{sec:3.3}).

\subsection{Lensed event rate}
\label{sec:rates}

\begin{table*}
\caption{Estimates for the total (lensed) GW event rates (per year) detected by different GW detectors with different SNR thresholds.}
\begin{adjustbox}{center}
\begin{threeparttable}

\renewcommand\arraystretch{1.1}
\begin{tabular}{cccccccc}		\toprule
Detector    &$\varrho_{0}$ & doub   & quad      & total &$\rm{quad}^{*}$  &$\rm{total}^{*}$ &   detectable          \\ \midrule
LIGO A+        &5    & $6.38 $$\times$$ 10^{-4} $ & $1.15$$ \times $$10^{-3} $ & $1.79 $$\times$$ 10^{-3} $ & $1.73 $$\times$$ 10^{-4} $ & $8.11 $$\times$$ 10^{-4} $  & $2.18$$\times$$10^2$ \\
LIGO A+        &8    & $4.97 $$\times$$ 10^{-5} $ & $1.81 $$\times$$ 10^{-4} $ & $2.31 $$\times$$ 10^{-4} $ & $1.02 $$\times $$10^{-5} $ & $5.99$$ \times$$ 10^{-5} $  & $5.37$$\times$$10^1$  \\
LIGO A+        &10   & $1.25$$ \times $$10^{-5} $ & $3.98 $$\times$$ 10^{-5} $ & $5.23 $$\times$$ 10^{-5} $ & $1.98 $$\times$$ 10^{-6} $ & $1.44 $$\times$$ 10^{-5} $  & $2.76$$\times$$10^1$  \\ \midrule
LIGO Voyager        &5    & $9.92 $$\times$$ 10^{-2} $ & $1.08$$ \times $$10^{-1} $ & $2.07 $$\times$$ 10^{-1} $ & $2.34 $$\times$$ 10^{-2} $ & $1.23 $$\times$$ 10^{-1} $  & $3.22$$\times$$10^3$ \\
LIGO Voyager        &8    & $7.85 $$\times$$ 10^{-3} $ & $1.10 $$\times$$ 10^{-2} $ & $1.88 $$\times$$ 10^{-2} $ & $2.00 $$\times $$10^{-3} $ & $9.84$$ \times$$ 10^{-3} $  & $7.87$$\times$$10^2$  \\
LIGO Voyager        &10   & $2.15$$ \times $$10^{-3} $ & $3.64 $$\times$$ 10^{-3} $ & $5.79 $$\times$$ 10^{-3} $ & $4.81 $$\times$$ 10^{-4} $ & $2.63 $$\times$$ 10^{-3} $  & $4.04$$\times$$10^2$  \\ \midrule
ET          &5        & $1.81 $$ \times$$ 10^{1} $  &  $5.46 $$ \times$$ 10^{0} $  & $2.36 $$ \times$$ 10^{1} $ & $2.60 $$ \times$$ 10^{0} $   & $2.07 $$ \times$$ 10^{1} $    & $1.10$$\times$$10^5$     \\
ET          &8        & $3.45 $$ \times$$ 10^{0} $   &  $1.87 $$ \times$$ 10^{0} $  &  $5.32 $$ \times$$ 10^{0} $  & $6.47$$ \times$$ 10^{-1} $    & $4.10 $$ \times$$ 10^{0} $ & $2.89$$\times$$10^4$    \\
ET          &10       & $1.30 $$ \times$$ 10^{0} $   & $9.03 $$ \times$$ 10^{-1} $   & $2.20 $$ \times$$ 10^{0} $ & $2.67 $$\times$$ 10^{-1} $    & $1.57 $$ \times$$ 10^{0} $    & $1.48$$\times$$10^4$  \\ \midrule

CE          &5        & $1.26 $$ \times$$ 10^{2} $  &  $1.32 $$ \times$$ 10^{1} $  & $1.39 $$ \times$$ 10^{2} $ & $1.02 $$ \times$$ 10^{1} $   & $1.36 $$ \times$$ 10^{2} $    & $5.21$$\times$$10^5$     \\
CE          &8        & $5.74 $$ \times$$ 10^{1} $   &  $9.85 $$ \times$$ 10^{0} $  &  $6.73 $$ \times$$ 10^{1} $  & $6.30$$ \times$$ 10^{0} $    & $6.37 $$ \times$$ 10^{1} $ & $2.97$$\times$$10^5$    \\
CE          &10       & $3.47 $$ \times$$ 10^{1} $   & $7.80 $$ \times$$ 10^{0} $   & $4.25 $$ \times$$ 10^{1} $ & $4.25 $$\times$$ 10^{0} $    & $3.90 $$ \times$$ 10^{1} $    & $1.95$$\times$$10^5$  \\ \midrule
GLOC          &5        & $1.86 $$ \times$$ 10^{2} $  &  $1.47 $$ \times$$ 10^{1} $  & $2.01 $$ \times$$ 10^{2} $ & $1.26 $$ \times$$ 10^{1} $   & $1.99 $$ \times$$ 10^{2} $    & $6.46$$\times$$10^5$     \\
GLOC          &8        & $1.05 $$ \times$$ 10^{2} $   &  $1.24 $$ \times$$ 10^{1} $  &  $1.17 $$ \times$$ 10^{2} $  & $9.23$$ \times$$ 10^{0} $    & $1.14 $$ \times$$ 10^{2} $ & $4.65$$\times$$10^5$    \\
GLOC          &10       & $7.21 $$ \times$$ 10^{1} $   & $1.08 $$ \times$$ 10^{1} $   & $8.29 $$ \times$$ 10^{1} $ & $7.39 $$\times$$ 10^{0} $    & $7.95 $$ \times$$ 10^{1} $    & $3.57$$\times$$10^5$  \\ \bottomrule
\end{tabular}
\begin{tablenotes}
\footnotesize
\item Notes: columns 1 and 2 list the name of the GW detector and the SNR threshold $\varrho_{0}$ set for the `detectable' GW events, respectively. Columns  3, 4, and 5 list the expected lensed GW event rates (per year) for those with double images (`doub'), quadruple images (`quad'), and both double and quadruple images (`total'), obtained by assuming the regular strategy introduced in Section~\ref{subsec:GWpEM}. Columns 6 and 7 with superscript notation `*' list the expected lensed GW event rates (per year) for those with quadruple images (`$\rm{quad}^{*}$') and both double and quadruple images (`$\rm{total}^{*}$'), obtained by assuming the conservative strategy introduced in Section~\ref{subsec:GWpEM}. The last column lists the expected detectable GW events without considering the gravitational lensing effect. 
\end{tablenotes}
\end{threeparttable}
\end{adjustbox}
\label{tab:1}
\end{table*}

We estimate the lensed event rates for both GW and EM signals from BNS mergers by assuming different conditions and different identification criteria. Table~\ref{tab:1} list the expected lensed event rates for the GW signals of BNS mergers that can be detected by different GW detectors, i.e., LIGO A+, LIGO Voyager, ET, CE, and GLOC, respectively. Apparently, ET, CE, and GLOC can detect $\sim 10^{4} - 10^{5}$ BNS mergers per year. Within this enormous amount of BNS mergers, it is promising to identify some lensed events by applying their waveform information. Adopting the regular criterion, for example, the total detectable lensed event rates of GW signals are expected to be $23.6$/$5.32$/$2.20$ per year for ET, and $139$/$67.3$/$42.5$ per year for CE if the SNR threshold is set as $\varrho_{0}=5/8/10$. Adopting the conservative criterion, the total detectable lensed event rates decrease to $20.7$/$4.10$/$1.57 $ per year for ET and $136$/$63.7$/$39.0$ per year for CE by setting $\varrho_{0}=5/8/10$, because less events with quadruple images rates can be detected.  The detectable event rate of BNS mergers is $218$/$53.7$/$27.6$ per year for LIGO A+ and $3220$/$787$/$404$ per year for LIGO Voyager corresponding to $\varrho_{0}=5/8/10$. However, the lensing probability for GW sources detected by LIGO A+ is $1$-$2$ order of magnitude lower than those by the other detectors, which leads to lensed event rate in the range of $10^{-3} - 10^{-5}{\rm yr^{-1}}$. This is mainly because the probability of sources being lensed by foreground galaxies is intrinsically lower at lower redshift. LIGO Voyager has a larger detection rate $\sim 10^{-1} - 10^{-3}{\rm yr^{-1}}$ than LIGO A+ does, which means that LIGO Voyager may be able to detect a lensed BNS within a period of ten years observation. GLOC, on the contrary, has the highest detection rate of the lensed BNS mergers, i.e., $201$/$117$/$82.9$ and $199$/$114$/$79.5$ per year, if adopting the regular and the conservative criterion by setting $\varrho_{0}=5/8/10$. 

\begin{figure*}
\includegraphics[width=\textwidth]{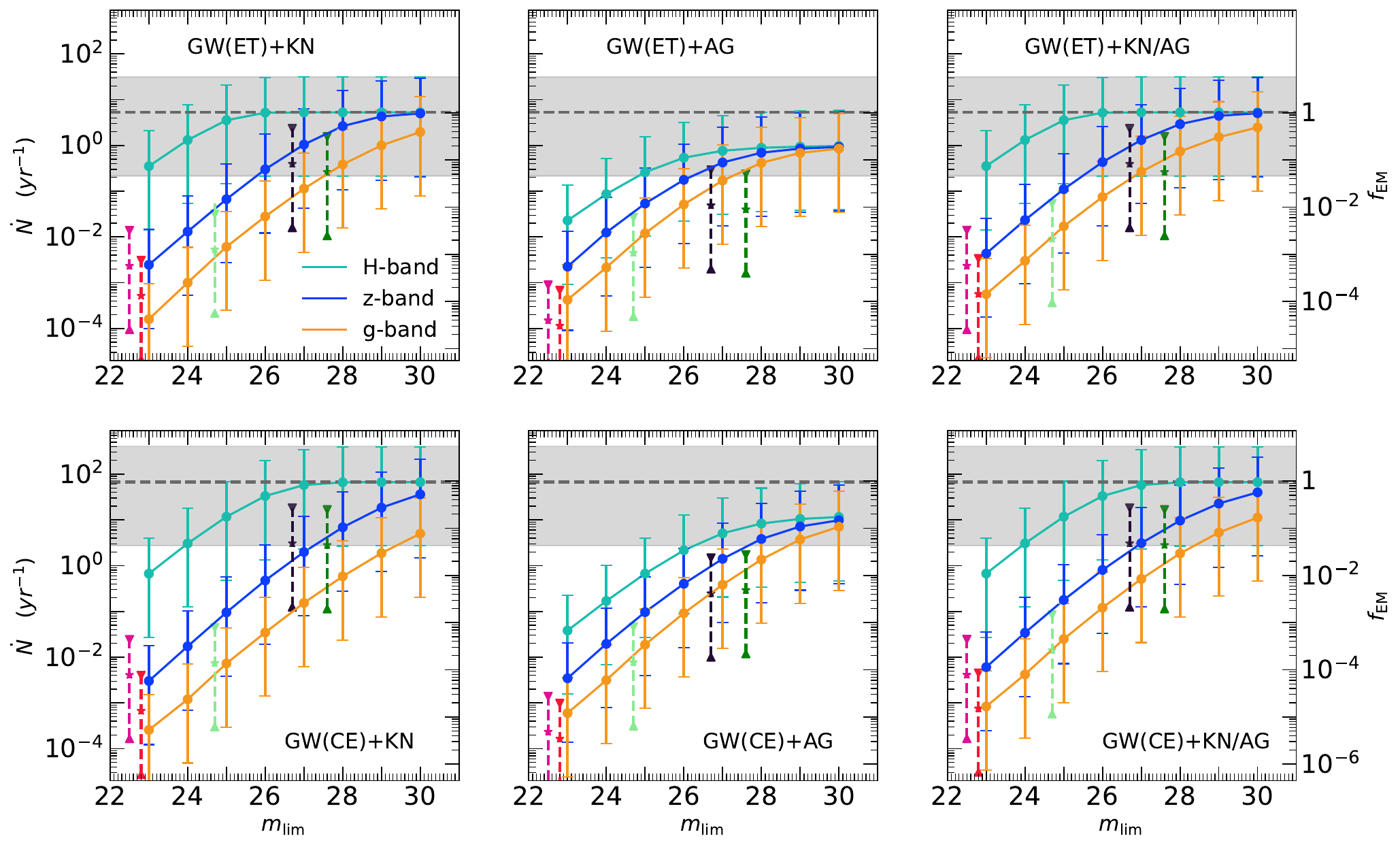}
\caption{
Estimates for the detectable GW+EM lensed event rate ($\dot{N}$) and the corresponding fraction of detectable EM counterparts among the lensed GW events from BNS mergers ($f_{\rm EM}$) as a function of the apparent limiting magnitude $m_{\rm lim}$. 
Top and bottom panels show the results obtained by using the ET and CE sensitivity curves, respectively. Solid lines and filled circles represent the expected all-sky detectable rates per year of the lensed events as a function of the limiting magnitude for telescopes at the g- (orange line), z- (blue line), and H-band (cyan line), respectively. For each panel, the left and right vertical axes mark the values of $\dot{N}$ and $f_{\rm EM}$, respectively. The star symbols stand for the estimate of the detectable EM event rates by considering the sky coverage and limiting magnitude for different telescopes. Magenta, red, green, black, and dark green star symbols from left to right represent those obtained for Euclid/H, 
Rubin/z, CSST/z, RST/H158, and JWST/F150W, respectively. The error bars associated with each filled circles and star symbols indicate the uncertainty of the estimate due to the uncertainty in the current constrained local merger rate density measured by the LIGO/Virgo detection. The horizontal dashed lines are the detectable lensed GW event rate listed in Table~\ref{tab:1}, for which the same uncertainty of local merger rate is given with shaded area. Left, middle, and right panels show the results obtained by considering the kilonova detection, the afterglow detection, and either the kilonova or afterglow detection, of those lensed GW events, respectively. The SNR threshold for the detection of GW signals is set as $\varrho_{0}=8$.
}
\label{fig:f4}
\end{figure*}

Figure~\ref{fig:f4} shows the expected detectable rate of the lensed GW and EM events as a function of the limiting magnitude of the telescope using the H-, z-, and g-band filters, respectively. We also correspondingly show the fraction of those with identifiable lensed hosts among the lensed GW events in the figure. Here we adopt the regular criterion to identify the lensed GW and EM events and only show the results for ET and CE for illustration. Apparently the rate of jointly detectable lensed GW+EM events increases with increasing limiting apparent magnitude of the adopted filter until it saturates. If the searching telescopes can reach sufficiently faint magnitude, all the kilonovae from the lensed GW events can be detected (see the cyan line in the left panel for H-band). However, only a fraction ($\sim 20\%$) of the afterglows associated with those lensed GW events can be detected (see the color lines in the middle panel) even if the limiting magnitude is very faint, which is mainly caused by that the afterglow emission is significantly anisotropic. Only those afterglows with a viewing angle less than the truncation angle $\theta_{\rm w}$ can be detected in the optical/infrared band, while kilonovae can be detected with any viewing angle. The rate for jointly detectable lensed GW+EM events by adopting a redder filter is larger than that by adopting a bluer filter with the same $m_{\rm lim}$. This is attributed to: 1) the intrinsic kilonova emission normally peaks at the optical band; 2) the afterglow intrinsically emits more in the redder band than in the bluer band at the rest frame time $\gtrsim 10^4$\,s \citep{Tolstov=2003}; and 3) the cosmological redshift effect causes the the optical emission shifting to the infrared band (as the redshift of lensed GW+EM events normally peaking around $z_{\rm s} \sim 1-2$ (see Section~\ref{sec:redmag}). 

Once a lensed BNS merger event is identified by a GW observatory, there are two ways to search for its EM counterparts: one is to search the whole localization area by a survey telescope with a given limiting magnitude; another is to directly observe the lensed host galaxy of the merger using a telescope with deep enough limiting magnitude if its host can be identified at prior by other surveys, which is mentioned above. Hereafter we only consider the second way, which is more efficient in searching for the EM counterparts. There are several factors that can affect the detection rate of the EM counterparts. First, only a fraction of the whole sky can be surveyed by a survey telescope. Second, the available sky region for a telescope at a given time may cover only a fraction of, or even not cover, the location of a transient phenomenon. Third, only a fraction of the lensed host galaxies can be identified by the survey telescope(s). Fourth, within the localization area given by GW detection lie many potential lensed host galaxies, which makes matching the lensed GW event and its host galaxy uncertain. Therefore, the rate for the EM detectable lensed events for a specific telescope should be down-scaled by these fractions, compared with the three colored lines shown in each panel of Figure~\ref{fig:f4}, i.e.,

\begin{equation}
\dot{N}_{\rm _{L,GW+EM,T}} = \dot{N}_{\rm _{L,GW+EM}} \cdot \frac{\Delta \Omega}{4 \pi} \cdot f_{\rm OL,T} \cdot f_{\rm Host} \cdot P_{\rm Host}.
\label{eq:GW+EM-T}
\end{equation}
Here $\Delta \Omega$ denotes the sky coverage of the adopted survey telescope, $f_{\rm OL,T}$ denotes the overlap ratio of the available sky region for the adopted searching telescope at any given time to the whole survey area for the lensed host galaxies, $f_{\rm Host}$ represents the fraction of the lensed host galaxies that can be identified by the adopted survey telescope, and $P_{\rm Host}$ represents the probability that the real lensed host galaxies can be well matched among the candidates within the localization of lensed GWs given by GW detection (success rate of matching).

For demonstration, we consider some current/future (survey) telescopes like Rubin, Euclid, CSST, RST, and JWST, respectively. Table~\ref{tab:2} lists the limiting magnitude, FoV, angular resolution, sky coverage and field of regard of Rubin, Euclid, CSST, RST, and JWST, respectively. The sky surveys taken by CSST and Euclid have similar sky coverage with $\Delta \Omega \sim 17,500$\,deg$^{2}$ and $15,000$\,deg$^{2}$, respectively (avoiding the light contamination from the Solar System and the Milky Way), and RST will perform a deeper survey but with a smaller sky coverage of $\Delta \Omega \sim  2000 $\,deg$^2$, overlapping with the survey area of CSST and Euclid. A large sample of lensed galaxies will be found by these surveys, among which some may be the hosts of BNS mergers. We assume the catalogs of lensed galaxies obtained by different surveys can all be used by the searching observations of the EM counterparts, thus we fix $\Delta \Omega=17,500$\,deg$^2$ for following rough estimates. The overlap ratio $f_{\rm OL,T}$ depends on the exact time of observations. For telescopes located at L2 (JWST, RST, and Euclid), the boresight pointing angle should remain in a specific range to avoid the sun light, within which the area within the circle along the ecliptic meridian is called the field of regard (FOR) and is $39\%$/$59\%$/$11\%$ for JWST/RST/Euclid\footnote{See more descriptions about FOR from \url{https://jwst-docs.stsci.edu/jwst-observatory-characteristics/jwst-observatory-coordinate-system-and-field-of-regard} for JWST, \url{https://roman.gsfc.nasa.gov/science/field_slew_and_roll.html} for RST, and \citet{Racca=2016} for Euclid.}. In general, the excluded sky survey area composed of ecliptic plane and the galactic plane is broadly uniform in the direction of ecliptic equator. Thus FOR are adopted as the overlap ratio $f_{\rm OL,T}$ for JWST, RST, and Euclid. As for Rubin, the overlap ratio $f_{\rm OL,T}$ is about 50\% \citep{chary=2019}. CSST has an orbit period about $90$ minutes and all sky can be observed in a short time, which means $f_{\rm OL,T} \sim 100\%$. As mentioned in Section~\ref{subsec:GWpEM}, for any given BNS merger GW event found in the sky area surveyed by these telescopes, the fraction of the lensed host galaxies that can be identified by the adopted survey telescope is fixed at $f_{\rm Host} \sim 30\%$. As for the last term, success rate of matching the lensed GW among the candidates in the potential sky area is also fixed at $P_{\rm Host}=100\%$.

\newcommand{\tabincell}[2]{\begin{tabular}{@{}#1@{}}#2\end{tabular}}
\begin{table*}
\centering
\caption{Properties of some current/future optical/infrared telescopes.}
\begin{adjustbox}{center}
\begin{threeparttable}
\renewcommand\arraystretch{1.1}
\begin{tabular}{lcccccc}
\toprule 
Telescope & filter & \tabincell{c}{$m_{\rm{lim}}$\\ $(\rm{Vega})$} & FoV & \tabincell{c}{resolution\\ (arcsec)}  & \tabincell{c}{sky coverage\\ $(\rm{deg^{2}})$} & \tabincell{c}{FOR\\ (at L2)}\\     \midrule
Rubin    & z     & 22.8 (30\,s) & 9.6\,deg$^{2}$ & 0.20 (0.70) & 18,000  &-\\
CSST   & z     & 24.7 (300\,s) & 1.1\,deg$^{2}$ & $0.15$  & 17,500  &- \\
Euclid & H     & 22.5 (336\,s) & 0.53\,deg$^{2}$ & $0.30 $ & 15,000  & 11\% \\
RST    & H158  & 26.7 (1\,hr) & 0.28\,deg$^{2}$ & $0.11$  & all-sky   & 59\% \\
JWST   & F150W & 27.6 (10\,ks) & 9.7\,arcmin$^{2}$ & $0.10$ & all-sky   & 39\% \\ 
\bottomrule
\end{tabular}
\begin{tablenotes}
\item Note: columns from 1 to 5 show the name of the telescope, the specific filter adopted in this paper, the limiting magnitude $m_{\rm lim}$ (in Vega magnitude), the field of view (FoV),the angular resolution (in arcsec) of the telescope, and the sky coverage (in unit of deg$^2$) of the (survey) telescope, field of regard (FOR), respectively. Note that all limiting magnitudes using AB magnitude are converted to Vega magnitude, RST and JWST can theoretically observe the all sky region for six months. For the ground-based Rubin telescope, the pixel pitch of the camera is 0.20\,arcsec/pixel, and we adopt a typical value of 0.70\,arcsec as its angular resolution considering median seeing conditions of the site. References: \citet{Ivezic=2019} for Rubin; \citet{Gong=2019} for CSST; \citet{Euclid2020} for Euclid; \citet{WFIRST2015} for RST; \url{https://jwst-docs.stsci.edu/jwst-near-infrared-imager-and-slitless-spectrograph} for JWST.
\end{tablenotes}
\end{threeparttable}
\end{adjustbox}
\label{tab:2}
\end{table*}

We estimate the detection rate of the EM counterparts for lensed BNS mergers by the CSST/Rubin/Euclid/RST/JWST-like telescope according to equation~\eqref{eq:GW+EM-T}. We find that CSST-like/Rubin-like/Euclid-like telescopes are unlikely to detect even a single lensed GW+EM event. CSST-like telescope appears to have the highest detection rate among these three, with $0.0141^{+0.0698}_{-0.0136} {\rm yr^{-1}}$ because of its relatively higher limiting magnitude $m_{\rm lim}$ and overlap ratio $f_{\rm OL,T}$. One may achieve higher limiting magnitude by increasing the exposure time for these three telescopes. If making the limiting magnitude $2$\,mag deeper than the current planned ones, it appears that Euclid and LSST can still hardly detect one such event even, though CSST may be able to detect one with several years observations (see Fig.~\ref{fig:f4}). RST-like and JWST-like telescopes, on the other hand, can detect $3.07^{+15.17}_{-2.95}$\,yr$^{-1}$ and $2.81^{+13.89}_{-2.70}$\,yr$^{-1}$ EM counterparts of the lensed BNS mergers identified by CE, and this number can be $0.39^{+1.94}_{-0.38}$\,yr$^{-1}$ and $0.26^{+1.29}_{-0.25}$\,yr$^{-1}$ for those identified by ET. It is promising that RST-/JWST-like telescopes can detect the EM counterparts for a few to several hundreds lensed BNS merger GW events within an observation period of ten years in the era of the 3rd generation GW detectors. Note here that if the success rate of matching the lensed GW among the candidates in the BNS localized area  is $P_{\rm Host}\sim 0.20$ \citep[][the pessimistic case]{Wempe=2022arXiv}, then the expected number of events that can be electromagnetically detected may be down scaled by a factor of $\sim 5$. Even in this case, our results will still suggest  that the EM counterparts of upto $\sim 6-36$ lensed BNS mergers can be detected within a detection period of ten years.

We adopt the simple kilonova model without considering the anisotropy of its emission, comparing with the anisotropic afterglow model. In reality, kilonova emission should be anisotropic and the observed magnitude depends on the viewing angle. Many factors, such as the geometry of kilonova, compositional inhomogeneity, and wavelength-dependent opacities, can contribute to this anisotropy, of which the efficacy may further depend on the spatial distribution and properties of the lanthanide-rich and lanthanide-poor ejecta material \citep{Perego=2017, Kasen=2017, Wollaeger=2018, Bulla=2019}. Some recent studies on the anisotropic kilonova found a factor of $\sim 2 - 3$ variation of the kilonova brightness at different viewing angles \citep{Grossman=2014,Perego=2014, Kasen=2015, Kasen=2017, Kawaguchi=2020, Darbha=2020}, which would introduce a variation of the peak magnitude over different viewing angles no larger than $1$\,mag. Our rough estimates show that the detectable lensed event rates may decline by a factor of $\lesssim 2$ for those telescopes with high limiting magnitude, e.g., JWST and RST, if assuming the peak magnitudes of all kilonovae are viewed at the angle with the faintest brightness, i.e., $1$\,mag lower than the simple model we adopted. This suggests that our conclusions will not be significantly changed if the anisotropic kilonova model is considered.

\begin{figure*}
\includegraphics[width=\textwidth]{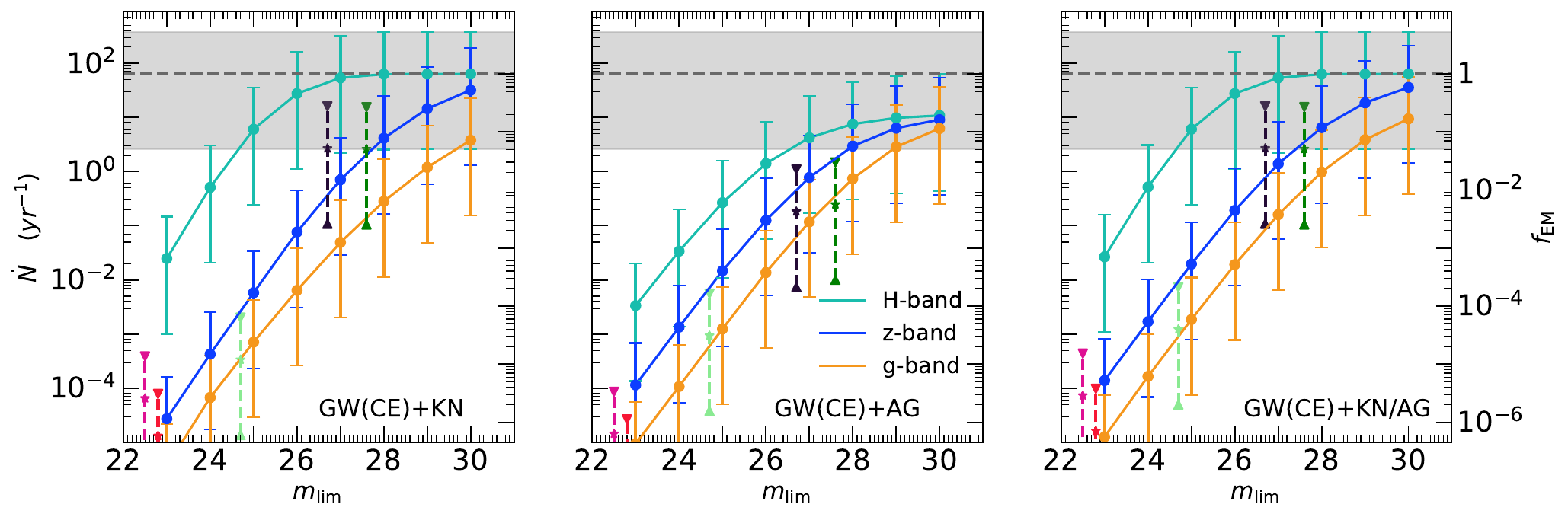}
\caption{
Similar to Figure~\ref{fig:f4} but adopting the conservative criterion, i.e., the faintest lensed images can be detected. In this figure, CE is adopted for the GW detection.
}
\label{fig:f5}
\end{figure*}

\begin{figure*}
\includegraphics[width=\textwidth]{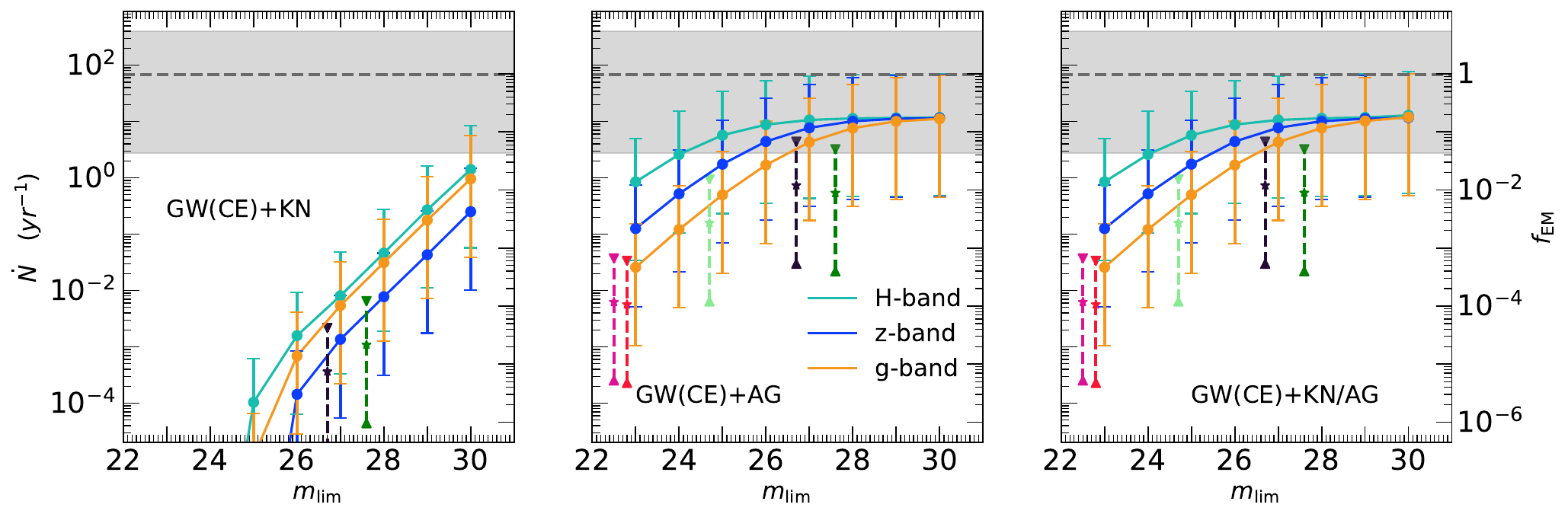}
\caption{
Similar to Figure~\ref{fig:f4} but the response time for afterglow searching is set to be $1$\,hr rather than $10$\,hr adopted in Fig.~\ref{fig:f4}. The magnitude of kilonovae are also set as the magnitude at $1$\,hr adopted in Figure~\ref{fig:f1}. In this figure, CE is adopted for the GW detection.
}
\label{fig:f6}
\end{figure*}

We also consider the conservative criterion, i.e., the faintest images of the lensed events can be detected by the joint GW+EM observations. Figure~\ref{fig:f5} shows the results on the expected rate for the detectable EM counterparts of the lensed BNS mergers detected by CE as a function of the limiting magnitude $m_{\rm{lim}}$ under the conservative criterion, similar to that shown in Figure~\ref{fig:f4} by adopting the regular criterion. Comparing with the bottom panels of Figure~\ref{fig:f4}, lensed event rate at lower limiting magnitude ($m_{\rm lim} \lesssim 25$) decline faster than that of higher limiting magnitude, which means that the quadruple-image case fraction increases as the limiting magnitude decrease. This is because, under the regular criterion, the magnification bias of quadruple-image cases is larger than that of double-image cases, since we apply the brightest image of quadruple-image cases (see also Figure~\ref{fig:f8}).

In addition, we also consider the cases that the response time of afterglow searching is shortened to $1$\,hr compared with $10$\,hr under the regular criterion. When the observing angle is small, the afterglow is substantially brighter if the response time can be short, and thus the detection rate of afterglows can be enhanced. The magnitude of kilonovae is also chosen as the magnitude at the phase of $1$\,hr, which is much ahead of the peak magnitude (see Figure~\ref{fig:f1}). Similarly as Figure~\ref{fig:f4}, Figure~\ref{fig:f6} shows the expected rate of the detectable EM counterparts for the lensed BNS mergers identified by CE as a function of the limiting magnitude $m_{\rm{lim}}$, except a quick response time of $1$\,hr is adopted. This change of response time enhances the detectability of afterglows, which enables the rate for the detectable afterglows to approach the upper limits at smaller $m_{\rm lim}$ (middle panel). Thus the chance of detecting a lensed event at lower limiting magnitude condition increases accordingly. Not surprisingly, detectable lensed event rate of kilonovae at the phase of $1$\,hr decline dramatically (left panel), which indicates that the advantage of afterglow detection over kilonovae detection at early phase. Note here that the g-band detection of kilonovae at the phase of $1$\,hr has higher rate over z-band detection, which is caused by the high energy emissions at the early phase of kilonova and can be also seen from Figure~\ref{fig:f1}.

Among those BNS mergers detected by GLOC, the number of lensed events increases slightly comparing with those by CE (see Table~\ref{tab:1}). Therefore, the expected rate for the detectable EM counterparts of the lensed BNS mergers identified by GLOC also increases accordingly. Since the number of lensed GW events of BNS mergers detected by LIGO A+/Voyager is small ($\lesssim$ $0.002$/$0.2$ per year), almost no lensed GW+EM signal from BNS merger is expected to be detected by it (see Table~\ref{tab:1}).

\subsection{Redshift and magnification}
\label{sec:redmag}
%

\begin{figure*}
\includegraphics[width=\textwidth]{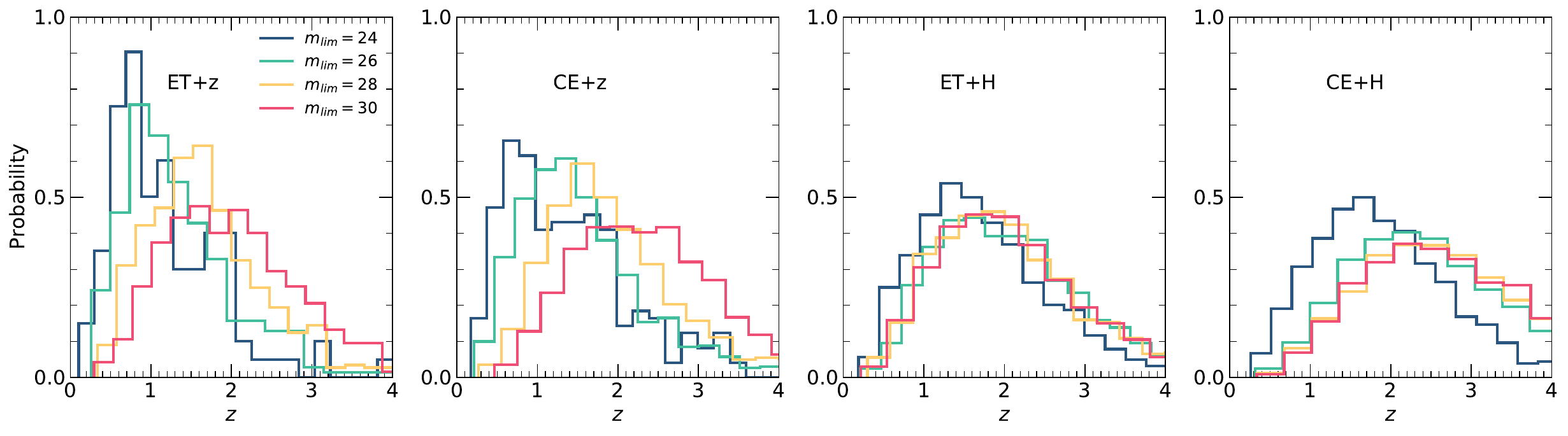}
\caption{
Redshift probability distributions of joint detectable lensed BNS merger events. Blue, green, orange, and red color histograms show the distributions for those lensed BNS mergers which are detectable by EM telescopes with the limiting magnitude of $24$, $26$, $28$, and $30$, respectively. Panels from left to right show the results for those events which can be detected by different combinations of GW detectors (ET, CE) and electromagnetic bandpass filters (z and H).
}
\label{fig:f7}
\end{figure*}

Figure~\ref{fig:f7} shows the redshift probability distributions of those lensed events detected by ET and CE that can be electromagnetically detected in the z- or H-band by an arbitrary survey with limiting magnitude of $24$, $26$, $28$, and $30$, respectively. As expected, the peak of the redshift distribution for GW+EM jointly detected events increases with increasing $m_{\rm lim}$. This peak is $z \sim 0.6-0.9$/$1.2-1.6$ for those lensed GW events detected by ET or CE and by an EM telescope with $m_{\rm lim}=24$ at the z-/H-band, while it shifts to $z\sim 2.0$ for those lensed GW events detected by ET or CE and by an EM telescope with $m_{\rm lim}=30$ at the z-/H-band. Furthermore, the redshift distribution is more extended for the cases with higher $m_{\rm lim}$, and those sources with redshift $z>3$ can be still detectable. These are simply caused by that the fainter $m_{\rm lim}$, the more distant events can be detected. Comparing the results of searching for the EM signals obtained by using the z-band with those by using the H-band,
as seen from Figure~\ref{fig:f7}, the H-band filter appears more efficient than the z-band filter in searching for the high redshift events, which illustrates that the redder filter band may be more suitable for searching lensed EM signals partly due to cosmological redshift effect. To be more specific, the spectra of the kilonova associated with GW170817 at 1.5 days peaks around $6,000$\AA\ in the optical band \citep{Pian=2017}. For such a kilonova at $z=1$, the peak of the spectra at the observer's rest frame  shifts to $12,000$\AA\ due to the cosmological redshift. For those lensed sources with even higher redshift, the cosmological redshift effect will be more pronounced. In addition, CE tends to detect many more high redshift events than ET does simply it is more sensitive than ET, especially at the low frequency end. 

\begin{figure*}
\includegraphics[width=\textwidth]{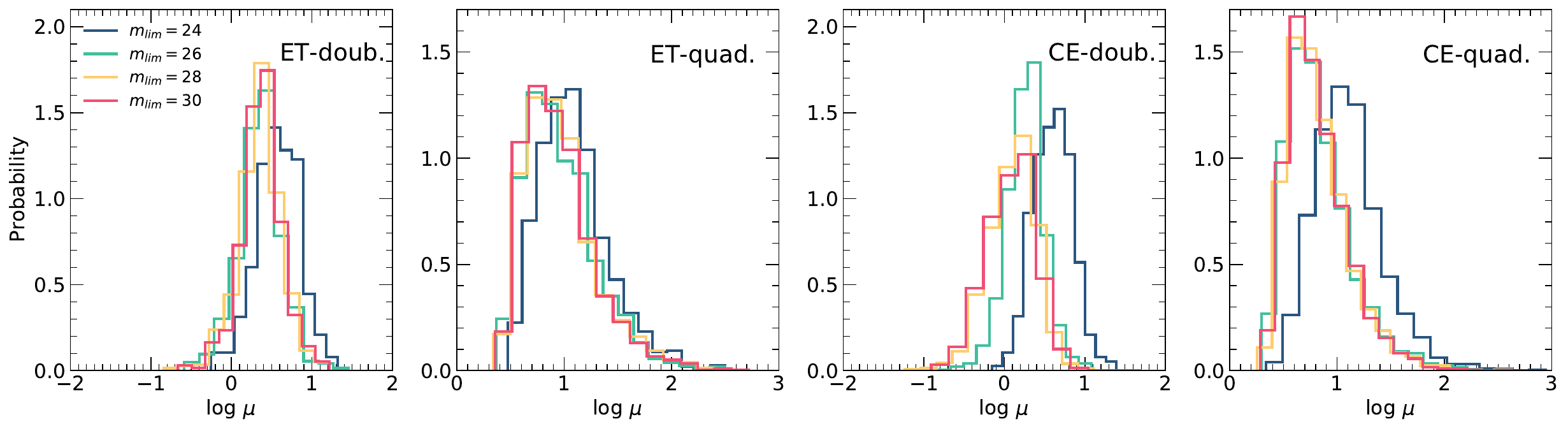}
\caption{Probability distributions of magnification factor $\mu$ of the second arrived images (the assumed first detected lensed EM images). The lensed BNS merger events are assumed to be detected electromagnetically in the H-band by a telescope with limiting magnitude of $24$ (blue), $26$ (green), $28$ (orange), and $30$ (red), respectively. Left two panels show the distributions for those BNS mergers detected by ET with double images (ET-doub) and quadruple images (ET-quad), and right two panels correspondingly for those detected by CE.
}
\label{fig:f8}
\end{figure*}

Figure~\ref{fig:f8} shows the probability distribution of the magnification factor ($\mu$) of the second arrived images for those GW and EM jointly detected lensed events. When $m_{\rm lim} \geq 26$ and lensed GW+EM event rates are nearly reaching the upper limit, the distribution of $\log \mu$ for double-image cases peaks around $0.2-0.5$ for both ET and CE (ET-doub and CE-doub panels), and the distribution of $\log \mu$ for quadruple-image cases peaks around $0.5-0.8$ for both ET and CE (ET-quad and CE-quad panels). However, when the limiting magnitude is comparably low ($m_{\rm lim}$ = 24), both doubly-imaged and quadruple-image cases prefer higher magnitude factor and $\log \mu$ peak around $0.5-0.7$ and $0.9-1.2$. This means that the resulting $\log \mu$ distribution for those jointly detected events does depend on the limiting magnitude $m_{\rm lim}$ set for searches of the EM counterparts, especially when $m_{\rm lim}$ is not high and the event rate is not approaching upper limit. We also note here that the $\log \mu$ distribution extends to $\log\mu <0$ ($\mu <1$) for double-image cases, which means that fainter images of double images may be de-magnified for some cases.

\subsection{Angular separations and time delays}
\label{sec:3.3}

\begin{figure}
\centering
\includegraphics[width=0.45\textwidth]{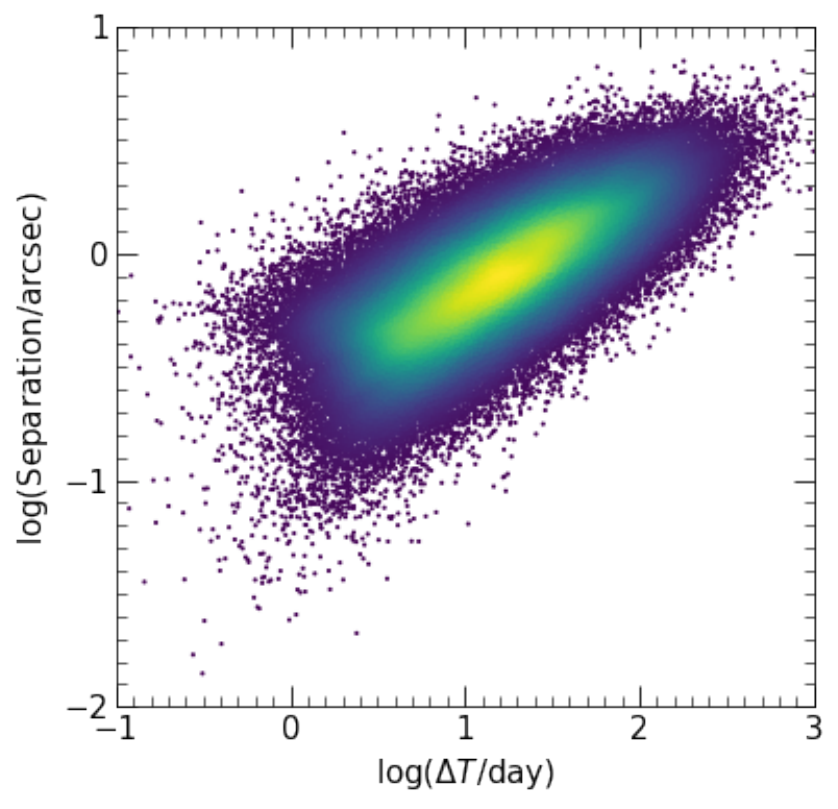}
\caption{
Distribution of the separations (in unit of arcsec) and time delays (in unit of day) for each image pairs obtained for those cases with double images. In this figure, each point represents a mock BNS merger lensed by an intervening galaxy, and totally $10^{5}$ mock BNS mergers are included in this density map (see details of our mock sample in Section~\ref{sec:3.3}).
}
\label{fig:f9}
\end{figure}

\begin{figure*}
\centering
\includegraphics[width=\textwidth]{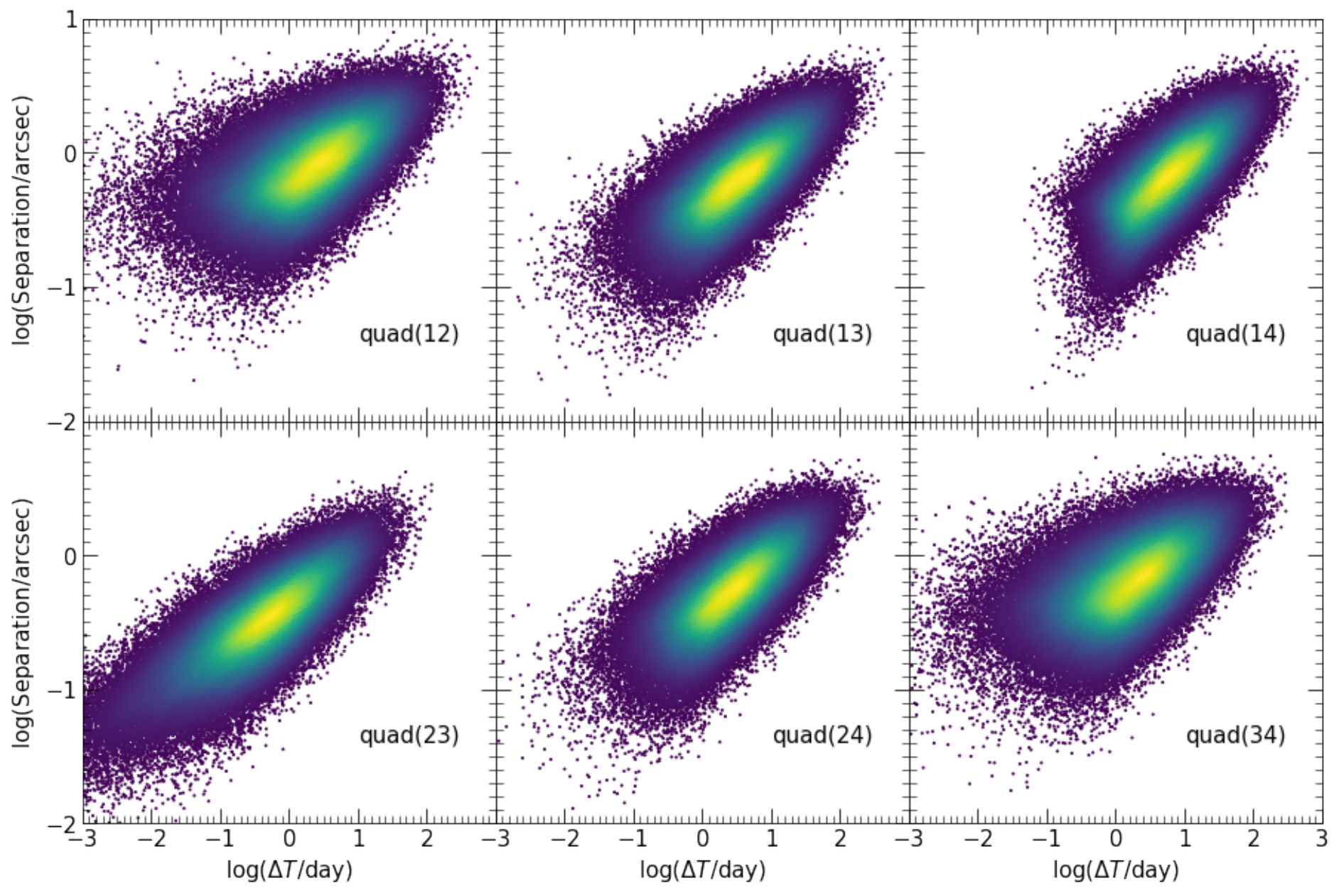}
\caption{Distributions of the separations (in unit of arcsec) and time delays (in unit of day) for each image pairs obtained for those cases with quadruple images. $10^{5}$ mock BNS mergers which are lensed by intervening galaxies are included in each of these density maps (see details of our mock sample in Section~\ref{sec:3.3}). The text label `quad($ij$)' represents the image pairs composed of the $i$-th and $j$-th images.
}
\label{fig:f10}
\end{figure*}

\begin{table*}
\label{tab:3}
\caption{Fractions of the lensed GW+EM events with different time delays and angular separations between different image pairs. Five telescopes are calculated assuming conservative strategy based on the GW detection of CE.}
\begin{adjustbox}{center}
\begin{threeparttable}
\renewcommand\arraystretch{1.1}
\begin{tabular}{clccccccc}		\toprule 
Image pairs& & $ {\rm doub}$ & ${\rm quad(12)}$ & ${\rm quad(13)}$ & ${\rm quad(14)}$ &${\rm quad(23)}$  &$ {\rm quad(24)}$ & ${\rm quad(34)}$  \\ \midrule
\multirow{5}{*}{Rubin} 
&$ P\left[\rm{Sep.}> 0.70\ ({\rm arcsec})\right]$  & 44.6\% &   46.4\% &  33.6\% & 36.7\% &   11.7\% & 26.1\% &   36.8\% \\ 
\cline{2-9}
&$ P\left[\Delta T > 10^{2}\ ({\rm s})\right]$ & $\sim 1$ &   98.6\% &  99.9\% & $\sim 1$ &   93.3\% & $\sim 1$ &   98.7\% \\
&$ P\left[\Delta T > 10^{3}\ ({\rm s})\right]$ & $\sim 1$ &   90.8\% &  97.1\% & $\sim 1$ &   70.2\% & 98.1\% &   90.5\% \\
&$ P\left[\Delta T > 1.0 \ ({\rm hr})\right]$  & 99.8\% &   77.4\% &  87.9\% & 98.3\% &   45.6\% & 86.6\% &   73.4\% \\
&$ P\left[\Delta T > 1.0 \ ({\rm day})\right]$ & 47.1\% &   13.9\% &  17.2\% & 25.7\% &    1.1\% &  9.6\% &    5.7\% \\  \midrule
\multirow{5}{*}{CSST} 
&$ P\left[\rm{Sep.}> 0.15\ ({\rm arcsec})\right]$    & 98.1\%  &   98.2\%  &  96.2\%  &   97.0\%  & 77.0\%  &  94.1\%  &  97.2\% \\
\cline{2-9}
&$ P\left[\Delta T > 10^{2}\ ({\rm s})\right]$ & $\sim 1$  &   99.4\%  &  $\sim 1$  &   $\sim 1$  & 95.7\%  &  $\sim 1$  &  99.3\% \\
&$ P\left[\Delta T > 10^{3}\ ({\rm s})\right]$ & $\sim 1$  &   96.9\%  &  99.5\%  &   $\sim 1$  & 80.1\%  &  98.5\%  &  93.4\% \\
&$ P\left[\Delta T > 1.0 \ ({\rm hr})\right]$  & $\sim 1$  &   90.8\%  &  96.6\%  &   99.9\%  & 60.4\%  &  92.3\%  &  82.1\% \\
&$ P\left[\Delta T > 1.0 \ ({\rm day})\right]$ & 71.9\%  &   28.7\%  &  35.1\%  &   49.0\%  &  5.1\%  &  24.1\%  &  17.0\% \\  \midrule
\multirow{5}{*}{Euclid} 
&$ P\left[\rm{Sep.}> 0.30\ ({\rm arcsec})\right]$    & 88.5\%  &    87.6\%  &  77.5\%  &   82.1\%  &   42.0\%  &   71.7\%  &  81.3\% \\
\cline{2-9}
&$ P\left[\Delta T > 10^{2}\ ({\rm s})\right]$  & $\sim 1$  &    99.7\%  &  $\sim 1$  &   $\sim 1$  &   93.6\%  &   99.9\%  &  99.3\% \\
&$ P\left[\Delta T > 10^{3}\ ({\rm s})\right]$  & $\sim 1$  &    95.2\%  &  99.1\%  &   99.9\%  &   77.0\%  &   98.5\%  &  93.9\% \\
&$ P\left[\Delta T > 1.0 \ ({\rm hr})\right]$   & 99.9\%  &    87.2\%  &  95.4\%  &   99.6\%  &   53.6\%  &   91.2\%  &  80.0\% \\
&$ P\left[\Delta T > 1.0 \ ({\rm day})\right]$  & 63.5\%  &    23.4\%  &  27.5\%  &   39.0\%  &    3.3\%  &   18.0\%  &  12.7\% \\  \midrule
\multirow{5}{*}{RST}
&$ P\left[\rm{Sep.}> 0.11\ ({\rm arcsec})\right]$   &  99.2\% &   99.3\% &  98.3\% & 98.8\% &   88.3\% &   97.7\% &   99.0\% \\
\cline{2-9}
&$ P\left[\Delta T > 10^{2}\ ({\rm s})\right]$ &  $\sim 1$ &   99.9\% &  $\sim 1$ & $\sim 1$ &   98.5\% &   $\sim 1$ &   99.8\% \\
&$ P\left[\Delta T > 10^{3}\ ({\rm s})\right]$ &  $\sim 1$ &   99.3\% &  99.9\% & $\sim 1$ &   93.5\% &   $\sim 1$ &   98.8\% \\
&$ P\left[\Delta T > 1.0 \ ({\rm hr})\right]$  &  $\sim 1$ &   97.0\% &  99.4\% & $\sim 1$ &   85.0\% &   99.3\% &   95.4\% \\
&$ P\left[\Delta T > 1.0 \ ({\rm day})\right]$ &  98.5\% &   62.6\% &  73.9\% & 87.4\% &   27.9\% &   64.8\% &   53.9\% \\  \midrule
\multirow{5}{*}{JWST} 
&$ P\left[\rm{Sep.}> 0.10\ ({\rm arcsec})\right]$    & 99.5\% &   99.7\% &  99.2\% & 99.8\% &   89.1\% &  98.9\% &  99.7\%  \\
\cline{2-9}
&$ P\left[\Delta T > 10^{2}\ ({\rm s})\right]$ & $\sim 1$ &   $\sim 1$ &  $\sim 1$ & $\sim 1$ &   98.5\% &  99.9\% &  99.8\%  \\
&$ P\left[\Delta T > 10^{3}\ ({\rm s})\right]$ & $\sim 1$ &   99.3\% &  99.9\% & $\sim 1$ &   94.2\% &  99.8\% &  98.6\%  \\
&$ P\left[\Delta T > 1.0 \ ({\rm hr})\right]$  & $\sim 1$ &   97.2\% &  99.5\% & $\sim 1$ &   86.5\% &  99.4\% &  96.3\%  \\
&$ P\left[\Delta T > 1.0 \ ({\rm day})\right]$   & 99.3\% &   62.3\% &  75.2\% & 90.7\% &   28.3\% &  70.6\% &  58.4\%  \\  \bottomrule
\end{tabular}
\label{tab:t3}
\begin{tablenotes}
\item Notes:  3, 4, 5, 6, 7, 8, and 9 columns (from top to bottom) list the fraction of time delays and spatial separation between the two image pairs for those cases with double images (denoted as `doub'), the second and first images (`quad(12)'), the third and first images (`quad(13)'), the fourth and first images (`quad(14)'), the third and second images (`quad(23)'), the fourth and second images (`quad(24)'), and the fourth and third images (`quad(34)'), of those cases with quadruple images, respectively.
\end{tablenotes}
\end{threeparttable}
\end{adjustbox}
\end{table*}

Figures~\ref{fig:f9} and \ref{fig:f10} show the two-dimensional distributions of angular separations and time delays for different image pairs of those mock lensed GW events with double and quadruple images, respectively. We obtain the mock samples by the following procedure. First, we sample $10^5$ detectable double-image cases and $10^5$ detectable quadruple-image cases with SNR $\varrho > 8$ following the same parameter distributions described in Section~\ref{sec:lensstat} and \ref{sec:gwlens}. Then, we solve the lens equations for each case to obtain the precise location for each lensed signals and calculate the time delays and angular separations between different lensed signals \citep[see detailed derivation in][]{Schneider=2006,LiSS=2018}. We also list in Table~\ref{tab:t3} the fraction of those lensed events, detectable by both CE and an EM telescope (Rubin/CSST/Euclid/RST/JWST) that have the time delays larger than $10^2$\,s, $10^3$\,s, $1$\,hr, and $1.0$\,day, and the fraction of those lensed events that with angular separation (between the two images of the double-image cases or between $i$-th and $j$-th arrived images of the quadruple-image cases) larger than the angular resolution of Rubin, CSST, Euclid,  RST, or JWST. All these fractions are obtained by similar procedures  as those introduced above but with additional EM conditions (limiting magnitude of each telescope, time delay cutoff, and angular separation cutoff), for which the EM counterparts parameter distributions adopted are the same as those in Section~\ref{sec:EMlens}.

We mainly consider the EM detection of the second arrived image but ignore the first image in this paper. The time delay of the second image to the first image ($\Delta T_{12}$) normally ranges from minutes to months as shown in Figures~\ref{fig:f9}, \ref{fig:f10} and Table~\ref{tab:t3}. In general, the kilonovae fade away with time at a rate of $0.3-2 \,\rm{mag}\, \rm{day^{-1}}$ after its peak luminosity \citep{Cowperthwaite=2017,Andreoni=2022}. For those afterglows with $\theta_{\rm obs} < \theta_w$, as we can see in Figure~\ref{fig:f3}, they also fade away with minor decline ($\lesssim 1\,\rm{mag}\, \rm{day^{-1}}$) in LCs. Thus for those cases with $\Delta T_{12}$ significantly larger than a few days, the EM signals of the first images may have already faded away (see Figure~\ref{fig:f1} and \ref{fig:f3}) and be hard to detect after the arrival of the second images. However, a small fraction ($\sim 1\%-50\%$) of the double-image cases have $\Delta T_{12} \lesssim 1$\,day (see Figure~\ref{fig:f10} and Table~\ref{tab:t3}), for which the EM signal of the first image can be still around its peak luminosity and thus can be also detected simultaneously with the EM signal of the second image. For those quadruple-image cases, the fraction of them having $\Delta T_{12} \lesssim 1$\,day becomes substantially larger (e.g., $\gtrsim 40\%$, see Figure~\ref{fig:f10} and Table~\ref{tab:t3}), and thus it may be easier to simultaneously detect the EM signals corresponding to the first images with the second ones. One may also similarly consider the simultaneous detection of the third and fourth EM images of those lensed GW events with quadruple images according to Figure~\ref{fig:f10}. 

We also note here that the time delay between different image pairs is also relevant to the GW detection of these lensed events. If the time delay between a image pair is shorter than the lifetime of the merger event observed by the GW detectors, the GW signals from both images may blend or interfere with each other, which leads to additional complexity in the data analysis. LIGO/Virgo can collect the GW signals of BNS mergers with a duration of $\sim 100$\,s, while future GW  detectors, i.e., ET and CE, may detect the BNS merger GW signals over a period of $\sim 1000$\,s, depending on the performance of their sensitivity at the low frequency end \citep{Zhao=2018,Chan=2018}. Almost all the cases with double images have $\Delta T_{12}>1000$\,s and thus the two GW signals can be well resolved in the time domain. For those lensed GW events with quadruple images, the situation is similar to the cases with double images, except that the time delay for the image pair `quad(23)' could be smaller than $1000$\,s for a relatively large fraction ($\sim 5\%-30\%$) of lensed events (see Table~\ref{tab:t3} and Figure~\ref{fig:f10}). Nevertheless, it should be safe to ignore the cases of GW signal overlap for the lensed BNS mergers.

The angular separation between a pair of lensed EM images must be large than the angular resolution of a telescope in order to resolve these two images if they are bright enough to be detected. As seen from Figures~\ref{fig:f9} and \ref{fig:f10}, the image pairs for most lensed GW events have angular separations larger than $0.1$\,arcsec and some of them ($36.2\%$ for those with double images, $5.9-37.6\%$ for those with quadruple images) have angular separation larger than $1$\,arcsec. Our results are broadly consistent with the angular separation for lensed quasars \citep{Yue=2022} and for lensed galaxies \citep{Collett=2015}. This suggests that the EM images of most EM and GW jointly detected lensed events can be resolved by spaceborne and ground-based telescopes with angular resolution as high as $\lesssim 0.1$\,arcsec, such as CSST, RST, JWST, and the next generation ground-based giant telescopes (like TMT, GMT and EELT) with adaptive optics if those EM signals are detectable, but those telescopes with angular resolution limited by seeing ($\gtrsim 0.7$\,arcsec; e.g., Rubin) may be difficult to resolve the lensed images separately. For those images that cannot be spatially resolved, the EM signals from different images may be blended with each other and thus lead to more complicated LCs. One possible method to overcome this is combining the time delays and LCs, by which LCs of the EM counterparts could show re-brightening signatures, through which those un-resolvable lensed events can be identified \citep{Chen=2021}.

\section{Conclusions and discussions}
\label{sec:conclusions}

In this work, we investigate the detectability of both the GW signals from lensed BNS mergers and their EM counterparts. We estimate the detectable rates of both the GW and EM signals from these mergers by future powerful GW detectors and telescopes by using the SIE lens model and simple models for the EM counterparts (afterglows and kilonovae) via the Monte Carlo method. Our main conclusions are summarized as follows.

\begin{itemize}
\item 
ET is expected to detect $5.32^{+26.1}_{-5.1}$ strongly lensed GW events from BNS mergers per year with at least two images having SNR larger than $8$, while CE stage-2 may detect $67.3^{+332}_{-64.7}$ such events per year. Among those lensed events, $4.10^{+20.2}_{-3.9}$ for ET and $63.7^{+315}_{-61.1}$ for CE stage-2 are expected to be detected with all the lensed images being detectable. LIGO Voyager may detect one within a period less than a decade. However, it is difficult for LIGO A+ to detect even a single lensed BNS merger within a period less than a decade.
\item The fraction of the detectable EM counterparts associated with those lensed GW events from BNS mergers that can be identified by a searching survey/telescope depends on the adopted filter and its limiting magnitude $m_{\rm lim}$. Adopting the ``redder'' band filters (e.g., H-band) can be more efficient in searching for the EM counterparts than adopting the ``bluer'' bands (e.g., z- and g-band). At $m_{\rm lim} \lesssim 26$, the detectable rate of H-band is at least one order of magnitude higher than that of z- or g-band given the same limiting magnitude condition. One should be cautious that this fraction estimation may be affected by some uncertainties, including those in the EM counterpart modelling, kilonova luminosity function. In addition, the response time of telescopes can also affect this fraction.
\item For realistic optical/infrared observation of those EM counterparts, we find that afterglow observation could be more optimistic than kilonovae observation in the first hour of telescope follow-up searches. On the contrary, kilonovae observation overwhelms in the later $ \sim 1-3$ days of observation.
\item RST-like and JWST-like telescopes can detect the EM counterparts of upto $3.07^{+15.17}_{-2.95}$\,yr$^{-1}$ and $2.81^{+13.89}_{-2.70}$\,yr$^{-1}$ lensed BNS merger GW events per year in the era of CE stage-2. With the observation over a period of several years or more, the cumulative number of such events will be sufficient for cosmological applications. However, Rubin-like, CSST-like, and Euclid-like telescopes can hardly have the chance to detect even a single event.
\item Redshift distributions of those lensed GW+EM events peak at $z \sim 0.6-0.9/1.2-1.6$ if they are detected by using the z-/H-band with $m_{\rm lim} = 24$\,mag, and the peaks shift to higher redshift ($z \sim 2$) if $m_{\rm lim}=30$\,mag. CE tends to detect more high redshift events than ET. In addition, the distribution of magnification factor $\log \mu$ of the second arrived lensed images peaks around $0.2-0.5/0.5-0.8$ for double-image/quadruple-image cases with limiting magnitude $m_{\rm lim} \gtrsim 26$. Adopting a lower limiting magnitude (e.g.,$m_{\rm lim} = 24$) for the searches, the resulting distributions peak around $0.5-0.7/0.9-1.2$ for double-image/quadruple-image cases.

\item Among those lensed GW+EM cases, $\sim 10\%-50\%$ of double-image cases and $\gtrsim 50\%$ of quadruple-image cases have $\Delta T_{12} \lesssim 1\,\rm{day}$. This means that the first arrived EM counterparts may still around their peak luminosity for those whose LC slowly decay (e.g., H-band LC of kilonova), and can be detected simultaneously with second arrived images. Similar consideration can also be taken for simultaneous detection of the third and fourth arrived images. 
Once we detect those multiple lensed image pairs, for RST and JWST that have angular resolution $\sim 0.1\,\rm{arcsec}$, most of the lensed events ($\gtrsim 85\%$) will be resolved except for the `quad(23)'.
\end{itemize}

The results obtained in this paper are instructive for future realistic searches of the lensed GW+EM events, which differ from those without considering lensing effect. One may have to take extra efforts for a successfully search of these lensed EM counterparts. First, it is required to quickly identify the lensed GW signals, i.e., the time between receiving two potential lensed GW alerts and identifying them as coming from the same lensed source needs to be as short as possible. Second, the lensed host galaxy signals also need to be quickly identified via all-sky galaxy surveys. One may compare the vague localization of the source by GW detectors (typically $\sim 1-10\,{\rm deg^2}$ for the third generation GW detector network) with the future complete deep all-sky survey of the lensed galaxies, and match the BNS merger with its lensed host galaxies through the observed time-delay and amplification information. Likewise, the time consumed by this step should be also as short as possible. Third, the response time of telescopes should not be too long so as not to miss those transients before they fade away.

We have investigated a number of factors that may affect the estimates for the rate of detectable GW+EM events, including the uncertainties in the BNS merger rate, different criteria set for the identification of those lensed events, different detectors and telescopes. However, there are other factors which may also affect the rate estimates. For example, we adopt simple models for kilonova and afterglow, for which a number of model parameters are fixed at values preferred by current observations of GW170817 and sGRBs (see Section~\ref{sec:EMlens}). In reality, each of these parameters may be different for different kilonovae and sGRBs. One may need to consider the distribution of these parameters to improve the estimates. In addition, the simplified kilonova model adopted in this paper is not taking the viewing angle and anisotropy into consideration. To consider the anisotropy of kilonova, it is required to specify the geometry, velocity, opacity, and other properties of both the dynamical ejecta and disk wind, which may be better understood by future NR simulations and constrained by observations. With many more detections of the kilonovae and afterglows in the future, we believe that both the kilonova and afterglow models can be improved better.

We note here that only single bandpass filters are considered to obtain the estimate for simplicity. One may need to further consider multiple bandpass filters for the searching of GW+EM counterpart by each specific (survey) telescope, which may also lead to some uncertainties in the rate estimates for the detectable GW+EM events. In addition, if the localization of some source can be determined precisely, it is reasonable to conduct EM searches in the given sky area with an exposure time longer and thus achieve a higher limiting magnitude than those listed in Table~\ref{tab:2}. Therefore, it may enable the detection of fainter images and leads to higher detectable lensed event rate. As stated before, a short response time of telescopes do not contribute to the detectability of kilonovae. However, a longer response time for GW trigger which may be weeks will decrease the detectability of both the kilonovae and afterglow depending on their decay rate of LCs.

Note also that the mission time overlap of different future GW detectors and EM telescopes is not considered for the estimates presented in this paper.
Rubin is about to coming into observation. JWST has started its observing and proved its powerful capability in the last few months. Euclid/CSST/RST are set to be launched in the next few year. LIGO will finish its upgrade to LIGO A+ by mid-2020s and then upgrade again to LIGO Voyager. ET and CE stage-1 may begin operation in 2030s, which may be at the twilight of the above EM telescopes designed lifetime. Possible extension of the mission time of these EM telescopes might overlap with the era of ET and CE stage-1, or even beyond. Considering this time line, it would be extremely lucky if a lensed GW+EM event could be detected in this decade, but it is possible to detect one or more lensed GW+EM events per year by ET/CE stage-1 and Euclid/RST/JWST-like telescopes in the next decade (2030s), and more than ten lensed GW+EM events per year in the era of CE stage-2 in 2040s. 

Furthermore, the next generation ground-based giant telescopes, i.e., Thirty Meter Telescope (TMT), Giant Magellan Telescope (GMT), and the European Extremely Large Telescope (EELT), are not considered in the present paper. These telescopes can achieve both deep field ($\sim27$\,mag) and high resolution ($\lesssim 0.1$\,arcsec) \citep{Michalowski=2021}, which may be also helpful in observing the EM counterparts of the lensed BN merger GW events.

\section*{Acknowledgements}
We thank the referee for helpful comments and suggestions. We thank Furen Deng for useful discussions. This work is partly supported by the National Natural Science Foundation of China (Grant Nos. 12273050, 11873056, 11690024, 11991052), the National Key Program for Science and Technology Research and Development (Grant No. 2020YFC2201400), and the Strategic Priority Program of the Chinese Academy of Sciences (Grant No. XDB 23040100). This research has made use of the SVO Filter Profile Service (http://svo2.cab.inta-csic.es/theory/fps/) supported from the Spanish MINECO through grant AYA2017-84089. 


\section*{Data Availability}

The data underlying this article will be shared on reasonable request to the corresponding author.

\UseRawInputEncoding
\bibliographystyle{mnras}
\bibliography{reference} 



%

\bsp 
\label{lastpage}
\end{document}